\documentstyle[prd,aps,twocolumn]{revtex}
\begin{document}
\draft

\twocolumn[\hsize\textwidth\columnwidth\hsize\csname @twocolumnfalse\endcsname
\title{QCD Confinement and $\Theta$ Vacuum: Dynamical Spontaneous Symmetry Breaking}
\author{Heui-Seol Roh \thanks{e-mail: hroh@nature.skku.ac.kr}}
\address{BK21 Physics Research Division, Department of Physics, Sung Kyun Kwan University, Suwon 440-746, Republic of Korea}
\date{\today}
\maketitle

\begin{abstract}
This study proposes that the longstanding problems of quantum chromodynamics (QCD) as
an $SU(3)_C$ gauge theory, the confinement mechanism and $\Theta$ vacuum, can be
resolved by dynamical spontaneous symmetry breaking (DSSB) through the condensation of
singlet gluons and quantum nucleardynamics (QND) as an $SU(2)_N \times U(1)_Z$ gauge
theory is produced.  The confinement mechanism is the result of massive gluons and the
Yukawa potential provides hadron formation. The evidences for the breaking of discrete
symmetries (C, P, T, CP) during DSSB appear explicitly: baryons and mesons without
their parity partners, the conservation of vector current and the partial conservation
of the axial vector current, the baryon asymmetry $\delta_B \simeq 10^{-10}$, and the
neutron electric dipole moment $\Theta \leq 10^{-9}$. Hadron mass generation mechanism
is suggested in terms of DSSB due to the $\Theta$ vacuum.
\end{abstract}

%\pacs{}
 ] \narrowtext

\section{Introduction}

Quantum chromodynamics (QCD) \cite{Frit} with quarks and gluons as fundamental
constituents is recognized as the fundamental dynamical theory for strong
interactions. One of the longstanding problems is however how to manage QCD in the low
energy region. The difficulty in treating QCD at low energy or long range comes from
the fact that the coupling constant becomes so strong that conventional perturbation
theory fails and confinement takes place in this limit so that free quarks and gluons
are not observed. Another longstanding problem of QCD is the $\Theta$ vacuum
\cite{Hoof2}, which is a superposition of the various false vacua, violating CP
symmetry.  This paper attempts to solve the problems nonperturbatively in terms of
dynamical spontaneous symmetry breaking (DSSB) from QCD, to demonstrate that quantum
nucleardynamics (QND) as an $SU(2)_N \times U(1)_Z$ gauge theory for nuclear
interactions originates from QCD as an $SU(3)_C$ gauge theory, and to propose that QND
for nuclear interactions is the analogous dynamics of the Glashow-Weinberg-Salam (GWS)
model as an $SU(2)_L \times U(1)_Y$ gauge theory \cite{Glas}. The DSSB mechanism is
adopted to strong interactions characterized by gauge invariance, physical vacuum
problem, and discrete symmetry breaking. The DSSB mechanism is different from the
Higgs mechanism in the GWS model, which has the problem in generating gauge boson mass
and fermion mass simultaneously. In this scheme, the only free parameter is the strong
coupling constant and several evidences for the violation of discrete symmetries
during DSSB are explicitly shown: baryons and mesons without their parity partners,
the conservation of vector current and the partial conservation of the axial vector
current, the baryon asymmetry $\delta_B \simeq 10^{-10}$ \cite{Stei0}, and the neutron
electric dipole moment $\Theta \leq 10^{-9}$ \cite{Alta}. Furthermore, the mechanism
of fermion mass generation and the quantization of intrinsic quantum number are
proposed as consequences of DSSB due to the $\Theta$ vacuum. This paper is restricted
to the real four dimensions without considering supersymmetry or higher dimensions.

QCD offers two remarkable characteristics, the asymptotic freedom
\cite{Gros} and confinement \cite{Wils0}. The discovery of the
asymptotic freedom in QCD leads to a convincing theory for strong
interactions; the renormalization group illustrates that the
running coupling constant at higher energies becomes smaller.
According to deep inelastic scattering experiments, the cross
sections show scaling invariance at higher energies, which means
that quark constituents behave like free particles at higher
energies, as illustrated by the renormalization group. Since the
running coupling constant becomes larger at lower energies
according to the renormalization group, it is suggested that
quarks and gluons are permanently confined inside a bound state.
However, a rigorous proof that quarks and gluons are confined does
not exist because conventional perturbation theory as an
appropriate formalism fails. This paper is thus intended to
introduce nonperturbative DSSB which demonstrates the confinement
mechanism rigorously as well as resolving the $\Theta$ vacuum
problem. Apart from technical difficulties associated with the
confinement mechanism, QCD has another problem known as the
$\Theta$ vacuum. Owing to the existence of non-trivial vacuum
gauge configurations, non-Abelian gauge theories have degenerate
vacuum configurations which are characterized by distinct homotopy
classes that can not be continuously rotated into one another. The
physical vacuum state of the theory is a superposition of all the
degenerate states. To resolve the $\Theta$ vacuum a
nonperturbative term, which affects neither the equations of
motion nor the perturbative aspects of the theory, may be added to
QCD Lagrangian density. The problem arises in this case from the
fact that nonperturbative effects violate CP, T, and P symmetries,
and would be relevant for an electric dipole moment for the
neutron unless the term is suppressed. The present experimental
bound to the electric dipole moment of the neutron, $d_n \leq
10^{-25} e \ \textup{cm}$, constrains the magnitude of the
nonperturbative term to be less than $10^{-9}$: it is known as the
strong CP problem \cite{Hoof2}.

In order to resolve the $\Theta$ vacuum or the strong CP problem, DSSB is adopted in
this scheme instead of introducing any dynamical fields such as axions
\cite{Pecc,Wein3}. QCD vacuum before DSSB is unstable as a local minimum energy state
and proceeds to a true vacuum which is a global minimum energy. DSSB due to the
condensation of singlet gluons triggers the axial current anomaly and makes the mass
reduction of color octet gluons. The massive gluon explains the confinement mechanism
represented by the Yukawa potential inside matter. The mechanism is accompanied by
massless gauge bosons known as Nambu-Goldstone (NG) bosons \cite{Namb}, which are
created by the mixing of gauge bosons. Convincing evidence for DSSB is the
existence of pseudoscalar scalar and vector mesons without their parity partners as
consequences of the violation of discrete symmetries during DSSB. The color vector
current is conserved but the color axial vector current is not conserved \cite{Adle}.
Therefore, this scheme simultaneously tries to resolve both the confinement mechanism
and the $\Theta$ vacuum in terms of DSSB.
The QCD confinement mechanism is thus linked to the general characteristics of QND as an
$SU(2)_N \times U(1)_Z$ gauge theory. The hadron mass is
generated as the DSSB of gauge symmetry and discrete symmetries, which is motivated by
the parameter $\Theta$ representing the surface term. Hadron mass generation
introduces constituent particle mass, dual Meissner effect, and hyperfine structure,
which justify the constituent quark model as the effective theory of QCD at low
energies. The $\Theta$ term plays important roles on the DSSB of the gauge group and
on the quantization of the hadron space and vacuum space. The $\Theta$ vacuum exhibits the
intrinsic principal number and intrinsic angular momentum for intrinsic space
quantization in analogy with the extrinsic principal number and extrinsic angular
momentum for extrinsic space quantization.

This paper is organized as follows. In Section II, QCD is reviewed to point out such
longstanding problems as the confinement mechanism and $\Theta$ vacuum. Section III
describes that DSSB by the condensation of singlet gluons makes the confinement
mechanism of QCD, in which massless gauge bosons are accompanied. In Section IV, QND
is addressed as the analogous dynamics of the GWS model for weak interactions. Section
V deals with hadron mass generation mechanism as the result of the breaking of gauge
and chiral symmetries. $\Theta$ constant and quantum numbers are discussed in Section
VI.    Section VII is devoted to conclusions.

\section{Review of QCD}

QCD for strong interactions is reviewed to outline the longstanding problems of the confinement mechanism
and $\Theta$ vacuum before proceeding to the resolution of the problems.

\subsection{$SU(3)_C$ Gauge Theory}

QCD as an $SU(3)_C$ gauge theory has three color charges and eight gluons;
the color singlet gluon is also taken into account in this scheme.
Natural units with $\hbar = c = k = 1$ are preferred for convenience throughout this paper unless otherwise specified.

The $SU(3)_C$ gauge-invariant Lagrangian density is, in four vector notation, given by
\begin{equation}
\label{qchr}
{\cal L } = - \frac{1}{2} Tr  G_{\mu \nu} G^{\mu \nu}
+ \sum_{i=1}  \bar \psi_i i \gamma^\mu D_\mu \psi_i + m \bar \psi_i \psi_i
\end{equation}
where the subscript $i$ stands for the classes of pointlike spinors, $\psi$ for the spinor and
$D_\mu = \partial_\mu - i g_s A_\mu$ for the covariant derivative with the coupling constant $g_s$.
Particles carry the local $SU(3)_C$ charges and the gauge fields are denoted by
$A_{\mu} = \sum_{a=0}^8 A^a_{\mu} \lambda^a /2$ with the Gell-Mann matrices $\lambda^a, a=0, . . , 8$.
Gell-Mann matrices satisfy the commutation relation
\begin{math}
[\lambda_k, \lambda_l ] = 2 i \sum_m f_{klm} \lambda_m
\end{math}
where $f_{klm}$ are the structure constants of the $SU(3)_C$ group.
The field strength tensor is indicated by
\begin{math}
\label{flst}
G_{\mu \nu} = \partial_\mu A_\nu - \partial_\nu A_\mu - i g_s [A_\mu, A_\nu] .
\end{math}
The fine structure constant $\alpha_s$ in QCD is defined such as the fine structure constant
$\alpha_e = e^2 /4 \pi$ in QED:
\begin{math}
\label{fins}
\alpha_s = g_s^2 / 4 \pi
\end{math}
where $\alpha_s$ is thus a dimensionless quantity.
The dimensionless coupling constant $g_s$ plays an important role in the renormalization of
gauge theory.

The set of unitary $3 \times 3$ matrices with $\textup{det} \ U =
1$ forms the group $SU(3)_C$ whose fundamental representation is a
triplet. The three color charges of a quark, $R, G$, and $B$
constitute the fundamental representation of the $SU(3)_C$
symmetry group. Gauge bosons for mesons are combinations of three
colors and three anti-colors which produce an octet and a singlet
gauge bosons; $3 \otimes \bar 3 = 8 \oplus 1$. The eight gluons in
the $SU(3)_C$ group theory are constructed as combinations of
colors, which are given by a matrix $\sum^8_1 \lambda_k A^k
/\sqrt{2}$:
\begin{equation}
\left ( \begin{array}{ccc}
A_3 + A_8/\sqrt{3} & R \bar G & R \bar B \\
G \bar R & - A_3 +  A_8/\sqrt{3} & G \bar B \\
B \bar R& B \bar G & - 2  A_8/\sqrt{3}
\end{array} \right )
\end{equation}
where two diagonal gluons are
\begin{eqnarray}
A_3 & = & (R \bar R - G \bar G)/\sqrt{2}, \nonumber \\ A_8 & = &
(R \bar R + G \bar G - 2 B \bar B )/ \sqrt{6} . \nonumber
\end{eqnarray}
The six off-diagonal gluons are accordingly represented by
\begin{eqnarray}
\label{glms} R \bar G & = & (A_1 - i A_2)/\sqrt{2}, \ R \bar B = (A_4 - i
A_5)/\sqrt{2},  \nonumber \\ G \bar B & = & (A_6 - i A_7)/\sqrt{2}, \ G \bar R  = (A_1
+ i A_2)/\sqrt{2},  \nonumber \\ B \bar R & = & (A_4 + i A_5)/\sqrt{2}, \ B \bar G =
(A_6 + i A_7)/\sqrt{2} .
\end{eqnarray}
The color singlet is symmetric:
\begin{equation}
A_0 = (R \bar R + G \bar G + B \bar B)/ \sqrt{3}
\end{equation}
which plays an important role in the confinement mechanism discussed in the following
section. It will be later realized that $A_0$ is the gluon with colorspin zero, $A_1
\sim A_3$ are degenerate gluons with colorspin one and $A_4 \sim A_8$ are
degenerate gluons with colorspin two:
the concept of colorspin will be introduced in the topic of intrinsic quantum
numbers. The gluon octet thus consists of three gluons with colorspin one and five
gluons with colorspin two.

\subsection{Phenomenology of QCD}

There are remarkable characteristics of QCD, the asymptotic freedom at higher energy,
the confinement at lower energy, and the $\Theta$ vacuum problem.
The confinement mechanism and $\Theta$ vacuum are briefly described before resolving them.

\subsubsection{Confinement Mechanism of QCD}

A notable characteristic is the confinement of quarks and gluons inside hadrons, but it seems not possible to
observe those individually since they are confined to the interior of hadrons \cite{Wils0}.
The confinement mechanism may be explained in terms of QCD but conventional perturbation theory is not applicable because
the behavior of the coupling constant becomes stronger at low energy region and calculation techniques beyond perturbation theory
are not reliable yet.
Even though the confinement mechanism is not perturbatively calculable in the strong coupling limit, nonperturbative interpretation
can be obtained in this paper as a rigorous proof of confinement.
This paper is intended to concentrate on this problem as well as the $\Theta$ vacuum.

\subsubsection{$\Theta$ Vacuum: Strong CP Problem}

QCD as a gauge theory has the $\Theta$ vacuum problem. The normal vacuum is unstable
and tunneling mechanism is possible between all possible vacua. The true vacuum must
be a superposition of the various vacua, each belonging to some different homotopy
class. The $\Theta$ vacuum can be therefore recast into a single, additional
nonperturbative term in the QCD Lagrangian density
\begin{equation}
\label{thet}
{\cal L}_{QCD} = {\cal L}_{P} + \bar \Theta \frac{g_s^2}{16 \pi^2} Tr G^{\mu \nu} \tilde G_{\mu \nu},
\end{equation}
\begin{equation}
\bar \Theta = \Theta + \textup{Arg} \ \textup{det} \ M_q
\end{equation}
where ${\cal L}_{P}$ is the perturbative part of the Lagrangian density (\ref{qchr}) except the explicit mass term, $G^{\mu \nu}$ is
the gluon field strength tensor,
$\tilde G_{\mu \nu}$ is the dual of the field strength tensor, and $M_q$ is the quark mass matrix.
Note that the effective $\bar \Theta$ term in the theory involves both the bare
$\Theta$ term relevant for pure QCD vacuum and the phase of the quark mass matrix relevant for electroweak effects.
Since the $G \tilde G$ term is a total derivative, it does not affect the perturbative aspects of the theory.
Such a term in the QCD Lagrangian violates CP, T, and P symmetries and is
related to the electric dipole moment of the neutron, $d_n \leq 10^{-25} e \ \textup{cm}$ \cite{Alta},
which constrains $\Theta$ to be less than $10^{-9}$: the $\Theta$ parameter is known as the strong CP problem.

As a topological term, instanton \cite{Hoof2} seems to provide a
plausible explanation for the fact that the massless gauge boson
associated with the breaking of the axial $U(1)$ symmetry is not
experimentally observed. It seems likely that the $\Theta$
structure of the QCD vacuum and the strong CP problem are to be
taken seriously since it solves the $U(1)_A$ problem. However,
instantons solve the $U(1)_A$ problem but raise the strong CP
problem. The idea by Peccei and Quinn \cite{Pecc} in order to
resolve the strong CP problem is to make $\Theta$ a dynamical
variable and is accomplished by introducing an additional global,
chiral symmetry: PQ symmetry. Weinberg et al. pointed out that
because the $U(1)_{PQ}$ is spontaneously broken global symmetry,
there must be a massless gauge boson, the axion, associated with
it \cite{Wein3}. In an axion model, the price for resolving the
strong CP problem is the existence of an additional, spontaneously
broken global symmetry and its associated pseudo-Goldstone boson:
the axion is however not observed. Several alternate models use
the idea of replacing the continuous symmetry by appropriate
discrete symmetries so that it leads to $\Theta = 0$ in the
Lagrangian \cite{Moha}. However, these models are not conclusive
in resolving the $\Theta$ vacuum problem.

In this proposal, DSSB is introduced to resolve the $\Theta$
vacuum problem. The bare $\Theta$ term in (\ref{thet}) added to
the $SU(3)_C$ Lagrangian density represents CP violation during
DSSB; the normal vacuum is full of massive gluons and leads to the
physical vacuum through the condensation of singlet gluons, which
causes the nonconservation of the axial vector current. This
resolves several unsolved problems in a single mechanism: the
$\Theta$ vacuum, the breaking of discrete symmetries, massless
gauge bosons associated with the $U(1)$ gauge symmetry, and the
confinement mechanism. The following section addresses DSSB to
resolve the $\Theta$ vacuum and confinement, and then the
resulting consequences are further concentrated on
phenomenological points of view in subsequent sections.

\section{Dynamical Spontaneous Symmetry Breaking: Confinement Mechanism}

In this section, the specific features of DSSB are focused on.
The color $SU(3)_C$ symmetry is dynamically spontaneously symmetry broken to the
$SU(2)_N \times U(1)_Z$ symmetry and then to the $U(1)_f$ symmetry; this is completely
analogous to the DSSB of the $SU(2)_L \times U(1)_Y$ symmetry to the $U(1)_e$ symmetry \cite{Glas}.
The combination of the confinement mechanism and $\Theta$ vacuum explains the DSSB mechanism in QCD analogous
to the Higgs mechanism in electroweak theory \cite{Glas,Higg}. The QCD Lagrangian
density (\ref{qchr}) possesses all the known strong interaction symmetries; it
conserves charge conjugation and parity. However, the physical vacuum is not
completely symmetric and DSSB from the normal to physical vacuum takes place. This
scheme uses dynamical symmetric breaking triggering the axial current anomaly
\cite{Adle} without introducing elementary scalar fields; it aims to have DSSB with
gauge interactions alone such as the motivation of technicolor models \cite{Suss}. The
concept of DSSB plays an important role in QCD which does not have essentially free
parameter. This proposal introduces color singlet gauge bosons as vacuum fields, which
also possess the $SU(3)$ symmetry and the dual property of discrete symmetries (P, C,
T, and CP) before DSSB.  However, DSSB takes places due to the surface effect, which
makes axial singlet gauge bosons among dual singlet bosons condense so as to violate
discrete symmetries. DSSB consists of two simultaneous mechanisms; the first mechanism
is the explicit symmetry breaking of gauge symmetry, which is represented by the color
factor $c_f$ and the strong coupling constant $g_s$, and the second mechanism is the
spontaneous symmetry breaking of gauge fields, which is represented by the
condensation of axial singlet gauge fields. The stable, physical vacuum is achieved
through DSSB due to the condensation of axial singlet gluons, which makes the
confinement for the resulting massive gluons. The condensation of axial singlet gauge
fields makes DSSB during which several characteristics appear: the dual Meissner
effect, discrete symmetry breaking, and confinement mechanism.

DSSB through the condensation of singlet gauge fields, color
dynamics, effective coupling constant and massive gluon, color
currents and color mixing angle, the breaking of discrete symmetries,
and massless gauge  bosons are focused on in the following.

\subsection{Dynamical Spontaneous Symmetry Breaking and $\Theta$ Vacuum}

In this scheme, the essential point is that color singlet gluons play roles of Higgs
particles in electroweak theory \cite{Glas,Higg}. Gauge fields are generally
decomposed by charge nonsinglet-singlet on the one hand and by even-odd discrete
symmetries on the other hand: they have dual properties in charge and discrete
symmetries. Color axial singlet gauge fields can be parameterized by scalar fields
representing vacuum fields regarded as the same $SU(3)$ symmetry. Singlet gauge fields
instead of scalar fields of the Higgs mechanism \cite{Higg} are thus used to produce
DSSB in this scheme.

Gluon interactions in the effective $SU(3)_C$ gauge invariant Lagrangian density with the bare $\Theta$ term are
\begin{equation}
\triangle {\cal L}^e = - \frac{1}{2} Tr  G_{\mu \nu} G^{\mu \nu} + \Theta
\frac{g_s^2}{16 \pi^2} Tr G^{\mu \nu} \tilde G_{\mu \nu} .
\end{equation}
Apart from charge nonsinglet gauge bosons, four singlet gauge boson interactions are
parameterized by the $SU(3)$ symmetric scalar potential:
\begin{equation}
\label{higs}
V_e (\phi) = V_0 + \mu^2 \phi^2 + \lambda \phi^4
\end{equation}
which is the typical potential with $\mu^2 < 0$ and $\lambda > 0$
for spontaneous symmetry breaking. The first term of the right
hand side corresponds to the vacuum energy density representing
the zero-point energy by non-axial singlet bosons. The axial
vacuum field $\phi$ is shifted by an invariant quantity $\langle
\phi \rangle$, which satisfies
\begin{equation}
\label{higs1}
\langle \phi \rangle^2 = \phi_0^2 + \phi_1^2 + \phi_2^2 + \phi_3^2
\end{equation}
with the condensation of the axial singlet gauge boson $\langle \phi \rangle = (\frac{- \mu^2}{2
\lambda})^{1/2}$. DSSB is connected with the surface term $\Theta \frac{g_s^2}{16
\pi^2} Tr G^{\mu \nu} \tilde G_{\mu \nu}$, which explicitly breaks down the $SU(3)_C$
gauge symmetry through the condensation of axial singlet gauge bosons and breaks
down the axial current. The $\Theta$ can be assigned by an dynamic parameter by
\begin{equation}
\label{thev}
\Theta = 10^{-61} \ \rho_G /\rho_m
\end{equation}
with the matter energy density $\rho_m$ and the vacuum energy density
$\rho_G = M_G^4$: the $\Theta$ constant is more discussed in
Section VI.

For mesons, DSSB takes place through the condensation of the color singlet gauge
field, $\langle \phi \rangle$, which induces the breakdown of the $SU(3)_C$ gauge symmetry. For
baryons, similar mechanism works for confinement; the color-color interaction rather
than the color-anticolor interaction occurs and creates an asymmetric triplet state
(color singlet) or a symmetric sextet state (color doublet). DSSB from QCD as an
$SU(3)_C$ gauge theory to QND as an $SU(2)_N \times U(1)_Z$ gauge theory is
accomplished through the condensation of axial singlet gluons with asymmetric color
combinations; the condensation makes gauge bosons less massive.

\subsection{Color Dynamics}

The behavior of gauge bosons depends on its energy and mass.
Gauge bosons, gluons, effectively behave like massless particles when their energies are higher than their masses
while they behave like massive particles when their energies are lower than their masses.
The color dynamics for gluons without considering their masses is firstly discussed \cite{Grif}
before the color dynamics for massive gluons with the lower energies is addressed.

\subsubsection{Formation of Meson: Quark and Antiquark Interactions}

The strong interaction amplitude is given by
\begin{equation}
{\cal M} = - \frac{g_s^2}{4} \frac{1}{k^2} [\bar u \gamma^\mu u][\bar v \gamma_\mu v]
(c_3^\dagger \lambda^a c_1)(c_2^\dagger \lambda_a c_4)
\end{equation}
where $u$ and $v$ denote quark flavor fields and $c_i$ with $i=1 \sim 4$ denote quark color fields.
The color factor is defined by
\begin{equation}
c_f = - \frac{1}{4} (c_3^\dagger \lambda^a c_1)(c_2^\dagger \lambda_a c_4) .
\end{equation}
The potential describing the quark-antiquark interaction is the same as the
Coulomb potential between two opposite charges
\begin{equation}
\label{copt}
V (r) = \frac{\sqrt{c_f \alpha_s}}{r}
\end{equation}
where the color factor depends on the color state of interacting quark.

Mesons are formed by quark and antiquark combinations:
in group theoretical language, $3 \otimes \bar 3 = 8 \oplus 1$
Quark-antiquark potentials are given by
\begin{equation}
V (r) = \sqrt{\frac{1}{6}} \frac{\sqrt{\alpha_s}}{r}
\end{equation}
for the color octet and
\begin{equation}
V (r) =  \sqrt{- \frac{4}{3}} \frac{\sqrt{\alpha_s}}{r}
\end{equation}
for the color singlet.
The positive color factor in the above represents the repulsion while the negative color factor represents the attraction.
This implies that repulsive forces between colored particles are dominant at
short range but attractive forces between color singlet particles are dominant at long range.
Note that the negative color factor leads to the imaginary potential relevant for damping.

\subsubsection{Formation of Baryon: Quark and Quark Interactions}

Baryons are combinations of three quarks.
Baryons are thus combinations of three color charges which produce a decuplet, two octets, and a singlet:
$3 \otimes 3 \otimes 3 = 10 \oplus 8 \oplus 8 \oplus 1$.

Phase transition from the $SU(3)_C$ symmetry to the $SU(2)_N \times U(1)_Z$ symmetry takes place in the formation of the baryon.
Two quark interactions between baryons, which is
$3 \otimes 3 = \bar 3 \oplus 6$ in group theoretical language, are the main channel for interactions.
The color factor defined by
\begin{equation}
c_f = \frac{1}{4} (c_3^\dagger \lambda^a c_1)(c_2^\dagger \lambda_a c_4)
\end{equation}
depends on the configuration of quarks and the potential is expressed by (\ref{copt}).
From two quarks, a triplet which has asymmetric combinations becomes
\begin{equation}
(RB - BR)/\sqrt{2}, \ (GB - BG)/\sqrt{2}, \ (GR - RG)/\sqrt{2}
\end{equation}
and a sextet which are symmetric combinations becomes
\begin{eqnarray}
&& RR, \ BB, \ GG, \nonumber \\ && (RB + BR)/\sqrt{2}, \ (GB +
BG)/\sqrt{2}, \ (GR + RG)/\sqrt{2} .
\end{eqnarray}
The color factor $c_f = 1/3$ for sextet configuration and $c_f = - 2/3$ for triplet configuration are obtained for quark-quark interactions.
For the color singlet baryon, each pair of quarks is
in a color $\bar 3$ since the pair must couple to the remaining quark to give a singlet.
Since six pairs with the same color factor are, together with the
normalization factor $1/6$, taken into account, the color factor is $- 2/3$.
Strong coupling constants for baryons are
$\alpha_b = c_f^b \alpha_s = \alpha_s/3 $, $\alpha_n = c_f^n \alpha_s = \alpha_s/4$,
$\alpha_z = c_f^z \alpha_s = \alpha_s/12$, and $\alpha_f = c_f^f \alpha_s = \alpha_s/16$
as symmetric color interactions and $- 2 \alpha_s/3 $, $- \alpha_s/2 $, $- \alpha_s/6$,
and $- \alpha_s/8$ as asymmetric color interactions.
The color factors are
$c^s_f = (c_f^b, c_f^n, c_f^z, c_f^f) = (1/3, 1/4, 1/12, 1/16)$
for symmetric interactions with even parity and $c^a_f = (-2/3, -1/2, -1/6, -1/8)$ for asymmetric interactions with odd parity;
the details are discussed in the following subsections.

\subsection{Effective Coupling Constant and Gluon Mass}

The previous subsection deals with gluons without considering their masses
but this subsection addresses effectively massive gluons.
In the absence of the gluon mass or in the case of the gluon with energy higher than mass ($M_G$), only the Coulomb potential exists in the
static limit ($E \rightarrow 0$) or ($E \rightarrow M_G$) respectively but in the presence of the gluon mass,
the Yukawa potential exists in the static limit ($E \rightarrow 0$).
Gluons with energies lower than their masses are thus taken into account so that
a rigorous proof for the confinement mechanism is shown in the following.

The strong interaction amplitude in the presence of the gluon mass is given by
\begin{equation}
\label{stia}
{\cal M} = - \frac{c_f g_s^2}{4} \frac{1}{k^2 - M_G^2} J_c^\mu J_{c \mu}^{\dagger}
\end{equation}
where the gluon mass $M_G$ is inserted in the gluon propagator. Parity or charge
conjugation violation due to the condensation of the singlet gluon must be taken into
account for current densities $J_c^\mu = [\bar u \gamma^\mu u] (c_3^\dagger \lambda^a
c_1)$ and $J_{c \mu}^{\dagger} = [\bar v \gamma_\mu v] (c_2^\dagger \lambda_a c_4)$.
That is, the color vector current is conserved (CVC) but the color axial vector
current is partially conserved (PCAC) just as the (V - A) current is conserved but the
(V + A) current is not conserved for weak interactions. The effective strong coupling
constant at low energy is expressed in analogy with the phenomenological, electroweak
coupling constant $G_F$:
\begin{equation}
\frac{G_R}{\sqrt{2}} = - \frac{c_f g_s^2}{8 (k^2 - M_G^2)} \simeq \frac{c_f g_s^2}{8 M_G^2}
\end{equation}
where $M_G$ indicates the mass of gluon, $k$ denotes the four momentum, and $c_f$
represents the color factor. Below critical temperature, the effective gauge boson
mass $M_G^{e} = (M_G^2 - E^2)^{1/2}$ increases as the gauge boson energy $E$ decreases
and at critical temperature, it reduces to zero.

Since the covariant derivative is changed from $D_\mu = \partial_\mu + i g_s A_{\mu a} \lambda^a/2$  to
$D_\mu = \partial_\mu + i g_n A_{\mu a}^{\pm} \lambda^a/2 + i g_z A_{\mu}^z$ in the Lagrangian density by phase transition,
the gauge boson mass term is obtained by
\begin{eqnarray}
\triangle {\cal L} & = & \frac{1}{2} (D_\mu A_{0})^2 - \frac{1}{2}
c_f^n g_s^2 \langle \phi \rangle^2 A_\mu^{\pm} A^\mu_{\pm} - \frac{1}{2} c_f^z
g_s^2 \langle \phi \rangle^2 A_\mu^z A^\mu_z  \nonumber \\ & = & \frac{g_n^2}{2}
(A_\mu^{\pm} A_{0})^2 - \frac{1}{2} g_n^2 \langle \phi \rangle^2 A_\mu^{\pm}
A^\mu_{\pm}   \cdot \cdot \cdot
\end{eqnarray}
where charged gauge fields $A_\mu^{\pm}$ and the longitudinal gauge field $A_\mu^z$
are defined in the next subsection and $\langle \phi \rangle$ is the condensation of the axial
singlet gauge boson. Note that the vacuum energy due to singlet gauge bosons $A_{0}$
is shifted with respect to the condensation $\langle \phi \rangle$; this is relevant for that the
condensation subtracts the zero-point energy in the region of QCD energy. The
essential point is that both the color coupling constant $c_f \alpha_s$ and the vacuum
expectation value $\langle \phi \rangle$ make the initially massive gauge boson lighter.
The gluon mass is thus generally reduced by the singlet gluon condensation $\langle \phi \rangle$:
\begin{equation}
\label{glma} M_G^2 = M_H^2 - c_f g_s^2 \langle \phi \rangle^2 = c_f g_s^2
[A_{0}^2 - \langle \phi \rangle^2]
\end{equation}
where $M_H = \sqrt{c_f} g_s A_{0}$ is the gauge boson mass at the grand unification
scale, $A_{0}$ is the singlet gauge boson, and $\langle \phi \rangle$ represents the condensation of
the axial singlet gauge boson. The color factor $c_f$ used in (\ref{glma}) becomes the
symmetric factor with even parity for singlet gauge boson and is the asymmetric factor
with odd parity for axial singlet gauge boson. The vacuum energy due to the zero-point
energy, represented by the gauge boson mass, is thus reduced by the decrease of the
color factor and the increase of the axial singlet gluon condensation as temperature
decreases. This process makes the breaking of discrete symmetries P, C, T, and CP,
which will be further discussed. The gluon mass is identical to the inverse of the
screening length $M_G = 1 /l_{QCD}$; the more the condensation the longer the
screening length. The gluon mass is thus the QCD cutoff scale $M_G = \Lambda_{QCD}$ at
the QCD phase transition. It is realized that the effective vacuum energy density
$V_e(\bar \phi)$ in (\ref{higs}) is related to the gauge boson mass $M_G$ by $V_e =
M_G^4$: $V_0 = M_H^4 \approx 10^8 \ \textup{GeV}^4, \ \mu^2 = - 2 c_f g_s^2 M_H^2
\approx - 10^5 \ \textup{GeV}^2, \lambda = c_f^2 g_s^4 \approx 36$ for the gluon mass
$M_G \approx 300$ MeV and the coupling constant $\alpha_s \approx 0.48$. The vacuum
energy density decreases in the order of $(\Lambda_{QCD}/M_H)^4 \approx 10^{-12}$ and
the system volume increases in the order of $(M_H/\Lambda_{QCD})^3 \approx 10^9$. The
QCD vacuum represented by massive gluons is quantized by the maximum wavevector mode
$N_R \approx 10^{30}$, the total gluon number $N_G = 4 \pi N_R^3/3 \approx 10^{91}$,
and the gluon number density $n_G = \Lambda_{QCD}^3 \approx 10^{-2} \ \textup{GeV}^3
\approx 10^{39} \ \textup{cm}^{-3}$.

The confinement for the color electric field can be illustrated more rigorously by considering the Yukawa potential \cite{Yuka} due to massive gluon.
The Yukawa potential associated with the massive gauge boson is given by
\begin{equation}
\label{yumu}
V (r) = \sqrt{\frac{c_f g_s^2}{4 \pi}} \frac{e^{-M_G (r - l_{QCD})}}{r}
\end{equation}
which shows the short range interaction for low energy gluons.

The essence in phase transition is that the massive gluon provides the short range force and the strength of the effective
coupling constant increases as the system expands.
The Coulomb potential proportional to $1/r$ at short range is transferred to the confinement
potential proportional to the Yukawa potential at long range.
Mesons are formed by the combination of quark and antiquark.
Baryons are regarded as three quark bound states:
$3 \otimes 3 \otimes 3 = 10 \oplus 8 \oplus 8 \oplus 1$.
Interactions between two quarks through color exchange can be expressed by the $SU(2)_N \times U(1)_Z$ gauge theory.
Therefore, the propagation of the massive gluon is limited in local place.
This is the reason why strong force phenomenologically gives the confinement of the color electric field
and is of short range: the dual Meissner effect.

\subsection{Color Currents and Color Mixing Angle}

Any choice of $\langle \phi \rangle$ which breaks a symmetry operator reduces
the mass for the octet gauge bosons such as shown in (\ref{glma}).
However, if the vacuum $\langle \phi \rangle$ is still left invariant by some
subgroup of gauge transformation, then gauge bosons associated
with this subgroup remain massless. The mixing of massive gauge
bosons of a gluon octet produces the color mixing angle
$\theta_R$, $\sin^2 \theta_R = 1/4$, so as to produce massless
gauge bosons. The color mixing angle is closely related to massive
gauge boson and massless gauge boson generation. The gauge mass
terms come from equation (\ref{glma}), evaluated at the shifted
vacuum $A_{0}^{'2} = A_{0}^2 - \langle \phi \rangle^2 $: in the case of
color-color interactions for baryons, gauge bosons involved are
$A_0 \sim A_8$. The relevant terms in the phase transition process
of the $SU(3)_C \rightarrow SU(2)_N \times U(1)_Z \rightarrow
U(1)_f$ symmetry are thus
\begin{equation}
A'^2_{0} (g_s A_\mu^a \lambda^a ) (g_s A^{\mu b} \lambda^b ) \\
\rightarrow A'^2_{0} [g_n^2 (A_\mu^{\pm})^2 + g_z^2 (A_\mu^z)^2 ]
\end{equation}
where $g_b^2 = c_f g_s^2$, $g_n = g_b \cos \theta_R$, and $g_z = g_b \sin \theta_R$ with
$\cos \theta_R = \frac{g_n}{g_b} = \frac{g_n}{\sqrt{g_n^2 + g_z^2}}$ and
$\sin \theta_R = \frac{g_z}{g_b} = \frac{g_z}{\sqrt{g_n^2 + g_z^2}}$.
There are massive bosons
\begin{eqnarray}
A_\mu^{\pm} & = & \frac{1}{\sqrt{2}} ( A_\mu^1 \mp i A_\mu^2) , \\
B^0_\mu & = & \cos \theta_R A_\mu^{3} - \sin \theta_R A_\mu^8 .
\end{eqnarray}
The masses are $M_A^2 = g_n^2 (A_{0}^2 - \langle \phi \rangle^2) = \frac{1}{4}
g_s^2 [A_{0}^2 - \langle \phi \rangle^2]$ and $M_B^2 = M_A^2 /cos^2 \theta_R =
c_f g_s^2  [A_{0}^2 - \langle \phi \rangle^2] = \frac{1}{3} g_s^2  [A_{0}^2 -
\langle \phi \rangle^2]$ with $M_H = g_n^2 A_{0}$ respectively: $A_\mu^+ = RG$
and $A_\mu^- = GR$ for color-color interactions according to
(\ref{glms}).

The fourth vector orthogonal to $B_\mu$ is identified as massless gauge boson:
\begin{equation}
\label{phot}
C_\mu = \sin \theta_R A_\mu^{3} + \cos \theta_R A_\mu^8
\end{equation}
with the mass $M_C = 0$.
Two gauge fields $B_\mu$ and $C_\mu$ are orthogonal
combinations of the gauge fields $A_\mu^{3}$ and $A_\mu^8$ with the mixing angle $\theta_R$.
The current for massless gauge boson dynamics is thus the combination of the color-neutral currents $J_\mu^{3}$ and $j_\mu^8$.
The generators of this scheme satisfy the relation \cite{Gell}
\begin{equation}
\label{baqu}
\hat Q_f = \hat C_{3} + \hat Z_c/2
\end{equation}
with the longitudinal component of the colorspin operator $\hat C_3$ and the hyper-color charge operator $\hat Z_c$
so that the corresponding current density is presented by
\begin{equation}
j_\mu^f = J_\mu^{3} + j_\mu^8/2 .
\end{equation}
The quantization of this scheme is resulted in the proton number conservation,
which is more discussed in the section of QND.
The interaction in the mixing color current can be given by
\begin{eqnarray}
\label{necu} -i g_n J_\mu^{3} A^{3 \mu} & - & i g_z j_\mu^8 A^{8
\mu}  \nonumber \\ & = & -i [g_n \sin \theta_R J_\mu^{3} + g_z
\cos \theta_R j_\mu^z/2 ] C^\mu \nonumber \\
& - & i [g_n \cos \theta_R J_\mu^{3} -
g_z \sin \theta_R j_\mu^z/2 ] B^\mu \nonumber \\ & = & - i g_f
j_\mu^f C^\mu - \frac{i g_f}{\sin \theta_R \ \cos \theta_R}
[J_\mu^{3} - \sin^2 \theta_R j_\mu^f] B^\mu \nonumber \\ & = & - i
g_f j_\mu^f C^\mu - i g_b [J_\mu^{3} - \sin^2 \theta_R j_\mu^f]
B^\mu
\end{eqnarray}
where the relation
\begin{equation}
g_f = g_n \sin \theta_R = g_z \cos \theta_R = g_b \cos \theta_R \sin \theta_R
\end{equation}
is used.
Therefore, fine structure constants $\alpha_b$ and $\alpha_f$ are due
to the exchange of the mixing $A_1 \sim A_3$ and $A_8$,
$\alpha_n$ is due to the exchange of the mixing $A_1$ and $A_2$,
and $\alpha_z$ is due to the exchange of the mixing of $A_3$ and $A_8$:
the coupling constant hierarchy is
$\alpha_b = c_f \alpha_s = \alpha_s /3, \ \alpha_n = \alpha_b \cos^2 \theta_R = \alpha_s/4$, $\alpha_z = \alpha_b \sin^2 \theta_R = \alpha_s/12$,
and $\alpha_f = \alpha_n \sin^2 \theta_R = \alpha_s/16$ with $\sin^2 \theta_R = 1/4$.
Gluons $A_4 \sim A_7$ with colorspin two are heavier than gluons $A_1 \sim A_3$ with colorspin one and
their contribution disappears during the phase transition of $SU(3)_C$ to $SU(2)_N \times U(1)_Z$ symmetry:
the detailed concept of colorspin will be discussed in the subject of intrinsic quantum numbers.
The condensation of singlet gluons results in the mixing of diagonal gluons $A_\mu^3$ and $A_\mu^8$ and
the coupling constant $\alpha_f$ is associated with the $U(1)_f$ gauge theory for
massless gauge boson dynamics.

\subsection{Breaking of Discrete Symmetries}

Strong interactions do not perturbatively violate isospin symmetry and are separately
invariant under parity inversion, charge conjugation, and time reversal. However,
discrete symmetries in strong interactions are nonperturbatively broken as seen in the
$\Theta$ vacuum problem. The $SU(3)_C$ gauge symmetry is spontaneously symmetry broken
to the $SU(2)_N \times U(1)_Z$ gauge theory and then to the $U(1)_f$ gauge theory by
the condensation of singlet gauge fields. The idea that singlet gauge fields condense is
similar to one in dual Landau-Ginzberg model approach to QCD \cite{Suzu}. The
condensation of singlet gauge fields makes the breakdown of discrete symmetries as
follows.

During phase transition, the discrete symmetries of time reversal (T), parity (P), and
charge conjugation (C) are violated so as to make hadrons massive: since the product
symmetry CPT remains intact, CP symmetry is violated. The breaking of discrete
symmetries through the condensation of singlet gluons is supported by looking at the
observation of pseudoscalar and vector mesons while their parity partners, scalar and
pseudovector mesons, are not observable; similarly, there is no baryon octet and
decuplet parity pairs. This resolves the $U(1)_A$ problem; the absence of the $U(1)_A$
particle is due to the nonconservation of the color axial vector current. This is
consistent with spin statistics for bosons that pseudoscalar meson is asymmetric in
color and isospin states and vector meson is symmetric in color and isospin states
since pseudoscalar meson is asymmetric in spin state and vector meson is symmetric in
spin state. This implies that singlet gauge bosons condense in the formation of the
hadron and the condensation is relevant for the partial conservation of axial vector
current (PCAC) $\partial_\mu J_\mu^5 = \frac{N_f c_f g_s^2}{16 \pi^2} Tr G^{\mu \nu}
\tilde G_{\mu \nu}$ with the flavor number $N_f$ and the conservation of vector
current (CVC) $\partial_\mu J_\mu = 0$: this is an example of parity violation in
strong interactions during phase transition. In the formation of hadrons, in other
words, the color doublet current is exactly conserved but the color singlet current is
not conserved. The $SU(3)_C$ symmetry is broken by DSSB, in which the $U(1)_A$
symmetry is not manifested in the particle spectrum and the Nambu-Goldstone boson
\cite{Namb} is absent as expected by the instanton mechanism \cite{Hoof2}. $C$
violation in baryon as three quark combination is shown in the number difference of
the proton and antiproton as observed in the baryon asymmetry of the present universe;
the baryon asymmetry requires C, T, and CP violation \cite{Sakh}.

Based on observation, there is also the tiny CP violation by the nonvanishing electric
dipole moment for the neutron $d_n = 2.7 \sim 5.2 \times 10 ^{- 16} \Theta \ e \
\textup{cm}$ \cite{Alta}. The experimental limit for an electric dipole moment of the
neutron implies that $\Theta \leq 10^{-9}$, which is known as the $\Theta$ vacuum
problem. The reason for such a small $\Theta$ is understood since, in the creation of
the neutron electric dipole moment, the condensation of singlet gluons takes place in
course of C and P violation together but the approximate conservation of CP produces
the small value of $\Theta$ during the confinement phase transition; CP violation is
minimal while both C violation and P violation are separately maximal during DSSB. CP
violation is small in the hadronic phase and this is compatible with the small
$\Theta$ value in the hadronic phase but it becomes bigger at higher energies. The
effective electric dipole length of the neutron becomes $l_d = G_R m_n$ and the
T-violation parameter is introduced by $\Theta = 10^{-61} \ \rho_G/\rho_m$ with
$\rho_m \simeq \rho_c \simeq 10^{-47} \ \textup{GeV}^4$. $\Theta \leq 10^{-9}$ at the
QCD cutoff energy and $\Theta \simeq 10^{-3}$ in the neutral kaon decay of weak
interactions. CVC or PCAC for color charges rather than flavor charges also indicates
DSSB with the massless gauge boson with the $U(1)_f$ gauge symmetry; the
nonconservation of the color axial vector current due to the $\Theta$ vacuum is
resolved by DSSB. The Cooper pairing between matter particles violates C symmetry and
the bare $\Theta$ term violates T symmetry explicitly. The phase of quark due to the
quark mass matrix distinct from the bare $\Theta$ vacuum is caused by isospin
interactions. Note that the extremely small $\Theta$ in strong interactions is
analogous to the small weak CP violation in weak interactions and is closely related
to the baryon asymmetry $\delta_B \simeq 10^{-10}$ in the present universe; the baryon
asymmetry is the consequence of the $U(1)_Z$ gauge theory, which will be discussed
more.  This thus gives the plausible resolution to the strong CP problem and observed
hadron bound states since DSSB solves the $U(1)_A$ problem and the strong CP problem
simultaneously.

The DSSB of gauge symmetry and chiral symmetry triggers the axial
current anomaly. This implies the reduction of zero modes via the
condensation of singlet gauge boson. The $\Theta$ vacuum as the
physical vacuum is achieved from the normal vacuum, which
possesses larger symmetry group than the physical vacuum.
The instanton mechanism as the vacuum tunneling is expected in the
Euclidean spacetime. The $\Theta$ vacuum term represents the
surface term since it is total derivative and decreases as the
system expands. Inflation, $(M_H/\Lambda_{QCD})^3 \approx 10^9$,
due to the gauge boson condensation therefore takes place during
phase transition since the effective vacuum energy density
represented by massive gauge bosons decreases as the vacuum
expectation value increases. The gluon mass is large in the small
system before phase transition while it is small in the large
system after phase transition.

\subsection{Massless Gauge Bosons}

This scheme proposes that massless gauge bosons (photons) as NG bosons \cite{Namb} are
created during the DSSB of QCD. The explicit examples of massless gauge bosons are the
intrinsic vibration modes in nuclear excitation with the typical energies $0.1 \sim
10$ MeV as noticed by the $\gamma$ decays. They are massless excited modes associated
with the generators of the $U(1)_f$ gauge symmetry. They are analogous to photons in
electroweak interactions as the  massless gauge modes indicating the breaking of gauge
symmetry. Massless gauge bosons are the quanta of the radiation field that describes
classical light. Phase transition from $SU(2)_N \times U(1)_Z$ to $U(1)_f$ gauge
symmetry produces massless photons with two transverse polarizations, which might stem
from the $SU(2)_N$ gauge symmetry. The $U(1)_f$ gauge theory for massless gauge boson
interactions is adopted as the analogy of the $U(1)_e$ gauge theory for photon
interactions. Massless gauge mode (photon) for the $U(1)_f$ gauge theory originated
from color charges has the coupling constant $\alpha_f = \alpha_s/16 = \alpha_n
\sin^\theta_R \simeq 1/34$, which is distinct from the coupling constant $\alpha_e =
\alpha_w \sin^\theta_W \simeq 1/137$ mediated by the photon for the $U(1)_e$ gauge
theory. However, the massless photon produced by the combination of color and isospin
charges has the coupling constant $\alpha_e \simeq 1/137$ at strong scale, which is
further discussed in the subject of QND; the conservation of the proton number is the
analogy of the conservation of the electron number. The analogy carriers over to a
correspondence between the theory of electromagnetic radiation in thermal equilibrium
and the theory of color radiation in thermal equilibrium. Each harmonic oscillator of
frequency $\omega$ can only have the energies $(n + 1/2) \omega$, where $n =0,1,2
\cdot \cdot$. This fact leads to the concept of massless gauge bosons as quanta of the
color field whose state is specified by the number $n$ for each of the oscillators
known as massive gluons.

Massless gauge bosons mediate the Coulomb potential ($\sim 1/r$)
and the condensation of singlet gluons makes the confinement
potential. As an analogy of the cosmic microwave background
radiation of phase transition at present universe, the clues of
phase transition for strong interactions remain as the  massless
gauge bosons; they have the Planckian form of spectra in the
region of the QCD cutoff energy. The average photon occupation
number in the thermal equilibrium is given by $f_{p} = 1 /(e^{E/T}
- 1)$. Massless photons are quantized by the maximum wavevector
mode $N_\gamma \approx 10^{29}$ and the total photon number $N_{t
\gamma} = 4 \pi N_\gamma^3/3 \approx 10^{88}$. The number density
of massless gauge bosons is given by $n_\gamma = 2 \zeta(3)
T^3/\pi^3 \approx 10^{29} \ \textup{cm}^{-3}$ at $ T = 1$ MeV. The
massless gauge boson is created by the nonconservation of the
axial current since it has odd parity. One thing for emphasizing
is that the photon might have mass due to the zero point energy
but it behaves like massless particle since its energy is always
greater than its mass. Note that nucleon excitation in the nuclear
system may be described by massless gauge boson interactions with
the predicted coupling constant $\alpha_f = \alpha_s/16$. The
$U(1)_f$ massless gauge boson dynamics might be useful in
describing the nuclear system as much as the $U(1)_e$ photon
dynamics. The excitation, alpha decay, and gamma decay of the
nuclear system are investigated by the $U(1)_f$ gauge theory,
which has the different origin with the electromagnetic
interactions involved by electrons.

\section{Analogies between Quantum Nucleardynamics and Glashow-Weinberg-Salam Model}

QCD leads to QND through DSSB as discussed in the previous section.
QND as an $SU(2)_N \times U(1)_Z$ for strong interactions is the analogous dynamics
of the GWS model as an $SU(2)_N \times U(1)_Z$ for weak interactions.
A number of problems have been
solved in terms of analogous property for the strongly interacting non-Abelian gauge theory.
In particular, the property
exhibits a specific mechanism of quark confinement; the weak
interaction due to isospin charges at higher energies is the
analogous description of the strong interaction due to colorspin
charges at lower energies. Analogies between strong force and
weak force are focused on from nonperturbative viewpoints.

There are several analogous properties between QND and GWS model. Both interactions
have characteristics of gauge theory, such as gauge invariance, vacuum problem, and
discrete symmetry breaking, and employ the DSSB mechanism in order to resolve unsolved
phenomena. Analogy exchanges strong coupling and weak coupling so that strong coupling
phenomenon can be analogously described by weak coupling phenomenon. The singlet weak
boson condensation is relevant for the nonconservation of the (V+A) current whereas
the singlet gluon condensation is relevant for the nonconservation of the axial vector
current. There are more analogies: colorspin charges and weak isospin charges, isospin
charge independence for QND and color charge independence for the GWS model, massless
gauge modes (photons) with the nuclear energy scale and massless gauge modes (photons)
with the electroweak energy scale, the vector current conservation (CVC) and (V - A)
current conservation, strong CP violation and weak CP violation, and the $\Theta$
vacua in strong and electroweak interactions. There are more conclusive clues for the
proposal that QND for strong force is the analogous dynamics of the GWS model for weak
force: fine structure constants $\alpha_n$ and $\alpha_w$, effective coupling
constants $G_R$ and $G_F$, and mixing angles $\theta_R$ and $\theta_W$ are more
focused on.

The fine structure constant $\alpha_s$ for strong interactions is measured by several experiments \cite{Hinc}:
\begin{math}
\alpha_s (M_Z) \simeq 0.12
\end{math}
at the momentum of the $Z$ boson mass $q = M_Z$
and
\begin{equation}
\alpha_s (q) \simeq 0.48
\end{equation}
at the momentum of nuclear energy $q \simeq 300$ MeV. Strong coupling constants for
baryons are described by $\alpha_b = c_f^b \alpha_s = \alpha_s/3$, $\alpha_n = c_f^n
\alpha_s = \alpha_s/4$, $\alpha_z = c_f^z \alpha_s = \alpha_s/12$, and $\alpha_f =
c_f^f \alpha_s = \alpha_s/16$ as symmetric color interactions and $- 2 \alpha_s/3 $,
$- \alpha_s/2 $, $- \alpha_s/6$, and $- \alpha_s/8$ as asymmetric color interactions:
$c_f^n = \sin^2 \theta_R$ and $c_f^f = \sin^4 \theta_R$. The color factors introduced
are $c^s_f = (c_f^b, c_f^n, c_f^z, c_f^f) = (1/3, 1/4, 1/12, 1/16)$ for symmetric
interactions and $c^a_f = (-2/3, -1/2, -1/6, -1/8)$ for asymmetric interactions. The
symmetric charge factors reflect intrinsic even parity with repulsive force while the
asymmetric charge factors reflect intrinsic odd parity with attractive force.
Asymmetric configuration for attractive force is confined inside particle while
symmetric configuration for repulsive force is appeared on scattering or decay
processes.

On the other hand, the fine structure constant $\alpha_w$ for the $SU(2)_L$ weak
interactions given by the data of muon decay \cite{Bardo} is
\begin{equation}
\alpha_w (M_W) \simeq 0.03
\end{equation}
at the momentum of the $W$ boson mass $q = M_W$. The other fine structure constants
for the GWS model are $\alpha_z = \alpha_w/\cos^2 \theta_W \simeq 0.04$ for $Z^0$
gauge boson exchange, $\alpha_y = \alpha_w \cos^2 \theta_W = \alpha_w/3 \simeq 0.01$
for a $U(1)_Y$ gauge theory, and $\alpha_e = \alpha_w \sin^2 \theta_W = \alpha_w/4
\simeq 1/133$ for a $U(1)_e$ gauge theory as symmetric isospin interactions at the
weak scale.

The color factors described above are pure color factors due to color charges but the
effective color factors used in nuclear dynamics must be
multiplied by the isospin factor $i_f^w = \sin^2 \theta_W = 1/4$ since the proton
and neutron are an isospin doublet as well as a color doublet.
Nucleons as spinors possess up and down colorspins as a
doublet just like up and down strong isospins:
\begin{equation}
{\uparrow \choose \downarrow}_c, \ \uparrow = {1 \choose 0}_c, \ \downarrow = {0
\choose 1}_c .
\end{equation}
This implies that conventional, global $SU(2)$ strong isospin symmetry introduced by Heisenberg \cite{Heis}
is postulated as the combination of local $SU(2)$ colorspin and local $SU(2)$ weak isospin symmetries;
this is further discussed in the subject of QND \cite{Roh31}.
Therefore, the effective color factors are given by
\begin{equation}
c_f^{eff} = i_f^w c_f = i_f^w (c_f^b, c_f^n, c_f^z, c_f^f) = (1/12, 1/16, 1/48, 1/64)
\end{equation}
for symmetric configurations. For example, the electromagnetic coupling constant for
the $U(1)_f$ gauge theory becomes $\alpha_f^{eff} = \alpha_s/64 \simeq 1/137$ when
$\alpha_s \simeq 0.48$ at the strong scale \cite{Hinc}; it is the same with the
electromagnetic coupling constant for the $U(1)_e$ gauge theory, $\alpha_e \simeq
1/137$.

The effective coupling constant for strong interactions $G_R$ and Fermi weak coupling
constant for weak interactions $G_F$ can be compared with each other. The ratio
between effective nuclear and weak forces $G_R/G_F = (\Lambda_{EW} / \Lambda_{QCD})^2
\approx 10^6$ is consistent with our expectation; the ratio between effective nuclear
and gravitational forces is $G_R/G_N = (\Lambda_{Pl} / \Lambda_{QCD})^2 \approx
10^{39}$. The gauge boson number density is given by $n_G = M_G^3$:  $n_{EW} \approx
10^{6} \ \textup{GeV}^3 \approx 10^{47} \ \textup{cm}^{-3}$ at the weak scale and
$n_{QCD} \approx 10^{-2} \ \textup{GeV}^3 \approx 10^{39} \ \textup{cm}^{-3}$ at the
strong scale.

The mixing angle for strong interactions $\sin^2 \theta_R = 1/4$ is the indication to
the DSSB of the $SU(2)_N \times U(1)_Z$ to the $U(1)_f$ gauge symmetry just as the
Weinberg angle for weak interactions $\theta^2_W = 1/4$ is the indication to the DSSB
of the $SU(2)_L \times U(1)_Y$ to the $U(1)_e$ gauge symmetry. These mixing angles
relate the strong coupling constant to the coupling for massless gauge boson dynamics,
$\alpha_f = \alpha_n \sin^2 \theta_R = \alpha_n/4$, and the weak coupling constant to
the coupling for photon dynamics, $\alpha_e = \alpha_w \sin^2 \theta_W = \alpha_w/4$,
respectively.

\section{Hadron Mass Generation}

In this section, hadron mass generation through DSSB triggered by the $\Theta$ vacuum
is described.
The DSSB mechanism is distinct from the conventional spontaneous symmety
breaking (SSB) mechanism or the Higgs mechanism \cite{Higg} since it is successful in generating the fermion
mass as well as the gauge boson mass.
The duality property before phase transition may be broken by DSSB after phase transition.

Conventional mass terms in the Lagrangian are not allowed because the left- and
right-handed components of the various fermion fields have different quantum numbers
and so simple mass terms violate gauge invariance. This scheme uses the DSSB of the
local $SU(3)_C$ gauge symmetry as well as global chiral symmetry \cite{Namb3}, which
uses the quark condensation as the order parameter, to generate hadron masses; the
DSSB mechanism of gauge symmetry and chiral symmetry is essential to give the hadron
mass. This mechanism is connected with the bare $\Theta$ vacuum term in (\ref{thet}),
which makes the nonconservation of the color axial singlet vector current rather than
the nonconservation of the flavor axial singlet vector current \cite{Pesk}. The
condensation of singlet gluons reducing the bare $\Theta$ vacuum generates hadron
masses as a consequence of gauge and chiral symmetry breaking in course of parity and
charge conjugation violation. The Lagrangian density of QCD except the $\Theta$ term
is symmetric but the physical vacuum of QCD is not chiral-symmetric; the QCD
Lagrangian density is invariant under the global $SU(3) \otimes SU(3)$ transformation
but the QCD vacuum is not invariant due to the condensation of singlet gluons and
quarks. A metastable, unphysical vacuum leads to a stable physical vacuum through DSSB
due to the condensation of singlet gluons. The $SU(2)_V$ color doublet vector current
is conserved but the $SU(2)_A$ (or $U(1)_A$) singlet color axial vector current is not
conserved during DSSB; this is analogous to electroweak interactions where the
$SU(2)_L$ doublet (V - A) current is conserved but the $SU(2)_R$ (or $U(1)_R$) singlet
(V + A) current is not conserved. Massless gauge bosons in the DSSB of gauge symmetry
and chiral symmetry appear as NG bosons. In the following, hadron mass generation is
addressed in terms of the DSSB of gauge symmetry and chiral symmetry, and is linked to
the hadron mass formula of the constituent quark model. The constituent quark mass is
relevant for the DSSB of strong interactions whereas current quark mass is relevant
for the DSSB of weak interactions; the generation of the constituent quark mass is
related to the bare $\Theta$ vacuum term whereas the generation of current quark mass
is related to the quark mass matrix term in (\ref{thet}).

Dual Meissner effect, constituent quarks, fine and hyperfine structure,
and hadron mass generation are addressed in the following.

\subsection{Dual Meissner Effect}

The hadron formation is the consequence of the color-color interaction due to the dual
Meissner effect, in which the color electric monopole and magnetic dipole (colorspin)
are confined inside the hadron while the color magnetic monopole and electric dipole
are confined in the vacuum. The difference number of axial-nonaxial singlet hadrons
$N_{sd} = N_{ss} - N_{sc}$ with the singlet number of constituent particles $N_{ss}$
and the condensation number $N_{sc}$ are introduced.

During the DSSB of gauge symmetry and chiral symmetry, the dual Meissner
effect of the color electric field in the relativistic case can be expressed by
\begin{equation}
\label{wame0}
\partial_\mu \partial^\mu A^\mu = - M_G^2 A^\mu
\end{equation}
where the right hand side is the screening current density,
$j^\mu_{sc} = - M_G^2 A^\mu$. Recall that the gluon mass is given
by $M_G^2 = M_H^2 - c_f g_s^2 \langle \phi \rangle^2 = c_f g_s^2 [A_{0}^2 -
\langle \phi \rangle^2]$ due to the DSSB of gauge symmetry breaking. The dual
Meissner effect of the color electric field in the static limit is
expressed by
\begin{equation}
\label{wame}
\nabla^2 \vec E_c = M_G^2 \vec E_c
\end{equation}
which exhibits the color electric field $\vec E_c$ excluded in the vacuum by
$\vec E_c = \vec E_{c0} e^{- M_G r}$ where $M_G = 1/l_{QCD}$.
Note the difference between the color dielectric due to the color electric
field $\vec E_c$ and the color diamagnetism due to the color magnetic field $\vec B_c$.
The mechanism is analogously connected with Faraday induction law, which opposes the change in the color electric flux rather than
the color magnetic flux, according to Lenz's law.

The gluon mass can be related to the hadron mass $m_h$:
\begin{equation}
\label{gafe}
M_G = (\frac{g_{sm}^2 |\psi(0)|^2}{m_h})^{1/2} \simeq \sqrt{\pi} m_h c_f \alpha_s
\sqrt{N_{sd}}
\end{equation}
or
\begin{equation}
\label{mase} m_h = \frac{g_{sm}^2 |\psi(0)|^2}{M_G^2} =
\frac{M_G}{\sqrt{\pi} c_f \alpha_s \sqrt{N_{sd}}}
\end{equation}
where $|\psi(0)|^2 = 1/l^3 = (m_h c_f \alpha_s)^3$ is the particle probability
density and $g_{sm} = 2 \pi n/\sqrt{c_f} g_s = 2 \pi \sqrt{N_{sd}}/\sqrt{c_f} g_s$ is the color magnetic coupling constant.
$N_{sd}$ is the axial-nonaxial singlet fermion
difference number in intrinsic two-space dimensions and $l$ is the macroscopic
length over which a wave function extends.
Equation (\ref{gafe}) is adopted from the analogy of electric superconductivity \cite{Aitc},
$M^2 = q^2 |\psi(0)|^2/m$: $q= - 2 e$ and $m = 2 m_e$ are replaced with
$g_{sm}$ and $m_h = \sum_i m_i$, respectively, where $m_i$ is the mass of each
constituent quark.

Fermion mass generation mechanism is the dual pairing mechanism of
constituent fermions, which makes paired fermions. According to
the electricity-magnetism duality \cite{Dira,Lee,Namb1,Seib}, the color
electric flux is quantized by $\Phi_E = \oint \vec E_c \cdot d
\vec A = \sqrt{c_f} g_s$ in the matter space while the color
magnetic flux is quantized by $\Phi_B = \oint \vec B_c \cdot d
\vec A = g_{sm}$ with the color magnetic coupling constant
$g_{sm}$ in the vacuum space: the Dirac quantization  condition
\begin{equation}
\sqrt{c_f} g_s g_{sm} = 2 \pi n = 2 \pi \sqrt{N_{sd}}
\end{equation}
is satisfied with the connection between $n$ and $N_{sd}$. In the
matter space, it is the pairing mechanism of color electric
monopoles while in the vacuum space, it is the pairing mechanism
of color magnetic monopoles according to the duality between
electricity and magnetism: color electric monopole pairing and
color magnetic monopole condensation \cite{Dira,Lee,Namb1,Seib}. In the
dual pairing mechanism, discrete symmetries P, C, T, and CP are
dynamically broken. Color electric monopole, color magnetic
dipole, and color electric quadrupole remain in the matter space
but color magnetic monopole, color electric dipole, and color
magnetic quadrupole condense in the vacuum space as the
consequence of P violation. Antimatter particles condense in the
vacuum space while matter particles remain in the matter space as
the consequence of C violation: the matter-antimatter asymmetry.
The electric dipole moment of the neutron and the decay of the
neutral kaon decay are the typical examples for T or CP violation.

A hadron system is effectively a collection of the Cooper pairs of
color doublets, so that the macroscopic occupancy of a single
quantum state can occur; in this state, all the pairs have the
same momentum in the center of mass frame. The non-zero value of
$N_{sd}$ implies the breaking of chiral symmetry. $N_{sd} = N_{ss}
- N_{sc}$ is the difference number of axial-nonaxial singlet
hadrons in intrinsic two-space dimensions, $N_{ss}$ is the singlet
hadron number, and $N_{sc}$ is the condensed hadron number.
Hadrons with the difference number $N_{sd}$ interact each other
with color symmetric configurations, $c^s_f = (c_f^b, c_f^n,
c_f^z, c_f^f) = (1/3, 1/4, 1/12, 1/16)$, while singlet hadrons
with the number $N_{ss}$ interact each other with color asymmetric
configurations, $c^a_f = (-2/3, -1/2, -1/6, -1/8)$. Note that
equation (\ref{mase}) is analogous to the pion decay constant
$F_\pi = - \langle \bar \psi \psi \rangle/2\langle \sigma
\rangle^2$ with the condensation of the sigma field $\langle
\sigma \rangle$ \cite{Suss}.

\subsection{Constituent Quarks}

Hadrons are postulated as composite particles consisted of constituent quarks according to the quark model, whose concept is
clarified in this part.
As the consequence of the $U(1)_Z$ gauge theory, the baryon number is conserved.

The relation between the gauge boson mass and the hadron mass, as confirmed by (\ref{mase}), is given by
\begin{math}
\label{gafe0}
M_H = \sqrt{\pi} m_h c_f \alpha_s \sqrt{N_{ss}}
\end{math}
or
\begin{equation}
\label{gafe5}
M_G = \sqrt{\pi} m_h c_f \alpha_s \sqrt{N_{sd}}
\end{equation}
where $N_{ss}$ is the number of singlet hadrons and $N_{sd}$ is the difference number
of axial-nonaxial singlet hadrons in intrinsic two-space dimensions. The hadron mass
formed as the result of confinement mechanism is composed of constituent quarks:
\begin{equation}
\label{gafe1}
m_h = \sum_i^N m_i
\end{equation}
where $m_i$ is the constituent quark mass.
In the above, $N$ depends on the intrinsic quantum number of constituent particles:
$N = N_{sd}^{3/2}$.
For examples, $N = 1/B$ with the baryon quantum number $B$ for the constituent quark in the formation of a baryon
and $N = 1/M$ with the meson quantum number $M$ for the constituent quark in the formation of a meson.

The difference number of fermions $N_{sd}$ is the origin of symmetry violation during
DSSB. Fermions with odd parity condense in the vacuum space while fermions with even
parity remain in the matter space; for example, magnetic monopoles with odd parity are
not observed but electric monopoles are observed in the matter space. Discrete
symmetries are violated so as to have complex scattering amplitude and the
nonconservation of the color axial singlet current. This is the main reason of the
change of the fermion mass and gauge boson mass.

\subsection{Fine and Hyperfine Structure}

In order to obtain mass formula, fine and hyperfine interactions are nonrelativistically considered to avoid complexity;
the static constituent quark model is justified.
Fine interactions are colorspin-colorspin interactions as the consequence of $SU(2)_N$ gauge theory and
hyperfine interactions include spin-spin and isospin-isospin interactions.

In QED, the dipole moment has the form expected for a Dirac pointlike fermion:
\begin{math}
\vec \mu_i = \frac{e}{2 m_e} \vec \sigma_i
\end{math}
where $e$ is the electric charge of particle, $m_e$ the mass of particle, and $\vec \sigma_i$ its Pauli matrix.
The spin-spin interaction due to the magnetic moment leads to the hyperfine splitting of the ground state:
\begin{equation}
\triangle E_{hf} = \frac{2}{3} \vec \mu_i \cdot \vec \mu_j |\psi (0)|^2
= \frac{2 \pi \alpha_e}{3} \frac{\vec \sigma_i \cdot \vec \sigma_j}{m_i m_j} |\psi (0)|^2
\end{equation}
where $\psi (0)$ is the wave function of two particle system at the origin $(r_{ij} = 0)$.
The above result can be taken over to colorspin-colorspin interactions as fine interactions in QCD:
\begin{eqnarray}
\label{fime}
\triangle E_{f} (\bar q q)
& = & \frac{2 \pi g_s^2}{3} \frac{\vec \zeta_i \cdot \vec \zeta_j}{m_i m_j} |\psi (0)|^2 , \\
\label{fiba}
\triangle E_{f} (q q)
& = & \frac{2 \pi g_s^2}{3} \frac{\vec \zeta_i \cdot \vec \zeta_j}{m_i m_j} |\psi (0)|^2 ,
\end{eqnarray}
where $g_s$ is the color coupling constant and $c_f$ is the asymmetric color factor.
The color factors are $c_f = \vec \zeta_i \cdot \vec \zeta_j = - 4/3$ for the color singlet of $\bar q q$
and $c_f = \vec \zeta_i \cdot \vec \zeta_j = - 2/3$ for the color triplet of $qq$.

The above result can be also extended to include spin-spin interactions:
\begin{eqnarray}
\label{hfme}
\triangle E_{hf} (\bar q q)
& = & \frac{2 \pi c_f g_s^2}{3} \frac{\vec \sigma_i \cdot \vec \sigma_j}{m_i m_j} |\psi (0)|^2 , \\
\label{hfba}
\triangle E_{hf} (q q)
& = & \frac{2 \pi c_f g_s^2}{3} \frac{\vec \sigma_i \cdot \vec \sigma_j}{m_i m_j} |\psi (0)|^2 ,
\end{eqnarray}
where $g_s$ is the color coupling constant and $c_f$ is the asymmetric color factor.
The color factors are $c_f = - 4/3$ for the color singlet of $\bar q q$ and $c_f = - 2/3$ for the color triplet of $qq$.
Isospin-isospin interactions are similarly
\begin{eqnarray}
\label{hfmt}
\triangle E_{hf} (\bar q q)
& = & \frac{2 \pi c_f g_s^2}{3} \sum_{i>j} \frac{\vec \tau_i \cdot \vec \tau_j}{m_i m_j} |\psi (0)|^2 , \\
\label{hfbt}
\triangle E_{hf} (q q)
& = & \frac{2 \pi c_f g_s^2}{3} \sum_{i>j} \frac{\vec \tau_i \cdot \vec \tau_j}{m_i m_j} |\psi (0)|^2 ,
\end{eqnarray}
where $\sum \vec \tau_i \cdot \vec \tau_j = 4 \vec i_i \cdot \vec i_j = 2 [i(i+1) - 3i(i+1)]$ with the total isospin $\vec I = \sum_j i_j$
and $i = 1/2$ for the proton and neutron.

\subsection{Hadron Mass Generation}

The hadron mass consists of three parts apart from the dual pairing mechanism:
constituent particle mass, fine structure energy, and hyperfine structure energy.

If fine structure contributions due to colorspin and isospin interactions are absorbed to the constituent quark mass,
the meson mass of the conventional constituent quark model is obtained:
\begin{equation}
\label{mema1}
m_m = m_1 + m_2 + A \frac{\vec \sigma_1 \cdot \vec \sigma_2}{m_1 m_2}
\end{equation}
where $\vec \sigma_1 \cdot \vec \sigma_2 = 4 \vec s_1 \cdot \vec s_2 =1$ for vector mesons and
$\vec \sigma_1 \cdot \vec \sigma_2 = -3$ for pseudoscalar mesons are given
and $A = \frac{8 \pi g_s^2 |\psi (0)|^2}{9}$.
The meson number of the constituent quark defined by $M = 1/2$ is not conserved in strong interactions.

Similarly, if fine structure contributions due to colorspin and isospin interactions are absorbed to the constituent quark mass,
the baryon mass of the conventional constituent quark model is obtained:
\begin{equation}
\label{bama1}
m_b = m_1 + m_2 + m_3 + A' \sum_{i>j} \frac{\vec \sigma_i \cdot \vec \sigma_j}{m_i m_j}
\end{equation}
where $A' = \frac{4 \pi g_s^2 |\psi (0)|^2}{9}$.
Since $\sum \sigma_i \cdot \sigma_j = 4 s_i \cdot s_j = 2 [s(s+1) - 3s(s+1)]$ with the total spin
$\vec S = \vec s_1 + \vec s_2 + \vec s_3$,
$\sum \vec \sigma_i \cdot \vec \sigma_j = 3$ for decuplet baryons and
$\sum \vec \sigma_i \cdot \vec \sigma_j = -3$ for octet baryons are given.
The baryon quantum number $B$ of a constituent quark defined by $B = 1/3$ is a conserved quantity.
For examples, the proton mass is given by
\begin{equation}
m_p = 3 m_u - A' \frac{3}{m_u^2} ,
\end{equation}
the $\Delta$ mass by
\begin{equation}
m_\Delta = 3 m_u + A' \frac{3}{m_u^2}  ,
\end{equation}
and the $\Sigma$ mass by
\begin{equation}
m_\Sigma = 2 m_u + m_s + A' [\frac{1}{m_u^2} - \frac{4}{m_u m_s}] .
\end{equation}

In this scheme, comparison between theoretical and experimental values shows remarkable agreement \cite{Rosn}.
This scheme thus justifies the constituent quark model \cite{Dash} as an effective model of QCD at low energies:
$m_u = m_d = 310$ MeV for mesons and $m_u = m_d = 363$ MeV for baryons.
Explicit numerical values are obtained in terms of the constituent quark model;
for meson, $\alpha_s = 0.48, \ |\psi (0)|^2 = (148 \ \textup{MeV})^3 , \ M_G = 300$ MeV, $A/m_u^2 = 94$ MeV
and for baryon, $\alpha_s = 0.48, \ |\psi (0)|^2 = (174 \ \textup{MeV})^3, \ M_G = 300$ MeV, $A'/m_u^2 = 55$ MeV.
For mesons, constituent strange quark has $m_s = 483$ MeV, $M_G = 300$ MeV, $\alpha_s = 0.48$, $|\psi (0)|^2 = (148 \ \textup{MeV})^3$ and
for baryons, constituent strange quark has $m_s = 536$ MeV, $M_G = 300$ MeV, $\alpha_s = 0.48$, $|\psi (0)|^2 = (174 \ \textup{MeV})^3$.
The effective coupling constant becomes $G_R \simeq 10 \ \textup{GeV}^{-2}$ for non-strange baryon interactions.
Note that the free parameter in QCD is only the coupling constant $\alpha_s$, in principle.
The mechanism of DSSB is in fact supported in other methods by the gluon condensation
$\langle G^a_{\mu \nu} G_{a \mu \nu} \rangle \approx 0.47 \ \textup{GeV}^4$ and
the quark condensation $\langle \bar u u \rangle = \langle \bar d d \rangle \simeq - (240 \ \textup{MeV})^3$ in the limit of zero quark masses.

In summary, masses of hadrons are results of local gauge and
global chiral symmetry breaking, which originate from the
condensation of singlet gluons and quark pairs. Hadron mass
generation due to this scheme is compatible with the constituent
quark mass as the consequence of the $U(1)_Z$ gauge theory. Due to
the DSSB of chiral symmetry, the vector current is conserved but
the axial vector current is partially conserved. The fine
structure of mass generation is contributed by the anomaly of
dipole-dipole interactions as the result of the $SU(2)_N$ gauge
theory. It suggests that hadron mass generation mechanism is
closely relevant for the axial current anomaly. The zero-point
energy is reduced by the DSSB of gauge and chiral symmetries. The
gauge boson mass $M_G$ is proportional to the hadron mass $m_h$
by $\sqrt{\pi} m_h c_f \alpha_s \sqrt{N_{sd}}$.

\section{$\Theta$ Constant and Quantum Numbers}

In the previous section, hadron mass generation triggered by the $\Theta$ vacuum is discussed and
in this section, intrinsic quantum numbers due to the $\Theta$ vacuum is addressed.

The $\Theta$ constant is parameterized by
\begin{math}
\Theta = 10^{-61} \ \rho_G/\rho_m
\end{math}
in (\ref{thev}).
This can be used for baryon mass generation from vacuum at the
strong scale:
\begin{equation}
\rho_B \simeq \Omega_B \rho_m \simeq 10^{-61} \Omega_B \rho_G/\Theta.
\end{equation}
If the conserved matter energy density in the universe is $\rho_m \simeq \rho_c \simeq 10^{-47} \ \textup{GeV}^4$,
the $\Theta$ constant in (\ref{thet}) depends on the gauge boson mass $M_G$ since $\rho_G = M_G^4$:
\begin{equation}
\label{thvu}
\Theta = 10^{-61} \ M_G^4/\rho_c \simeq 10^{-12}
\end{equation}
at the strong scale $M_G \simeq 10^{-1}$ GeV.
This result is consistent with the measured value $\Theta \leq 10^{-10}$ \cite{Alta} from
the neutron electric dipole moment.

\subsection{Intrinsic Quantum Numbers}

There is the condensation process in hadron mass generation mechanism. The process is
the dual pairing mechanism of singlet constituent fermions, which makes bosonlike particles
of paired fermions. At the phase transition, $N_{sc}$ becomes zero so that $N_{sd}$
becomes the maximum. Using relations $M_G = \sqrt{\pi} m_h c_f \alpha_s \sqrt{N_{sd}}$ and $M^2_G =
M^2_{H} - c_f g_s^2 \langle \phi \rangle^2 = c_f g_s^2 [A_{0}^2 - \langle \phi \rangle^2]$, the zero point energy
$M_{H}^2 = \pi m_h^2 c_f^2 \alpha_s^2 N_{ss}$ and the reduction of
the zero-point energy $\langle \phi \rangle^2 = m_h^2 c_f \alpha_s N_{sc}/4$ are obtained.
The difference number of axial-nonaxial singlet hadrons $N_{sd}$
in intrinsic two-space dimensions suggests the introduction of a degenerated
particle number $N_{sp}$ in the intrinsic radial coordinate and an intrinsic principal
number $n_m$; particle quantum numbers are connected with the relation $n_m^4 =
N_{sp}^2 = N_{sd}$ and the Dirac quantization condition
$\sqrt{c_f} g_s g_{sm} = 2 \pi N_{sp}$ is satisfied.
The $N_{sp}$ is thus the degenerated state number in the intrinsic
radial coordinate that has the same principal number $n_m$. The intrinsic principal
quantum number $n_m$ consists of three quantum numbers, that is, $n_m = (n_c, n_i,
n_s)$ where $n_c$ is the intrinsic principal quantum number for the color space, $n_i$
is the intrinsic principal quantum number for the isospin space, $n_s$ is the
intrinsic principal quantum number for the spin space. Intrinsic quantum numbers
$(n_c, n_i, n_s)$ take integer numbers. A fermion therefore possesses a set of
intrinsic quantum numbers $(n_c, n_i, n_s)$ to represent its intrinsic quantum states.

The concept automatically adopts the three types of intrinsic angular momentum
operators, $\hat C$, $\hat I$, and $\hat S$, when intrinsic potentials for color,
isospin, and spin charges are central so that they depend on the intrinsic radial
distance: for instance, the color potential in strong interactions is dependent on the
radial distance. The intrinsic spin operator $\hat S$ has a magnitude square $\langle
S^2 \rangle = s (s + 1)$ and $s = 0, 1/2, 1, 3/2 \cdot \cdot \cdot (n_s-1)$. The third
component of $\hat S$, $\hat S_z$, has half integer or integer quantum number in the
range of $- s \sim s$ with the degeneracy $2s + 1$. The intrinsic isospin operator
$\hat I$ analogously has a magnitude square $\langle I^2 \rangle = i (i + 1)$ and $i =
0, 1/2, 1, 3/2 \cdot \cdot \cdot (n_i-1)$. The third component of $\hat I$, $\hat
I_z$, has half integer or integer quantum number in the range of $- i \sim i$ with the
degeneracy $2i + 1$. The intrinsic color operator $\hat C$ analogously has a magnitude
square $\langle C^2 \rangle = c (c + 1)$ and $c = 0, 1/2, 1, 3/2 \cdot \cdot \cdot
(n_c-1)$. The third component of $\hat C$, $\hat C_z$, has half integer or integer
quantum number in the range of $- c \sim c$ with the degeneracy $2c + 1$. The
principal number $n_m$ in intrinsic space quantization is very much analogous to the
principal number $n$ in extrinsic space quantization and the intrinsic angular momenta
are analogous to the extrinsic angular momentum so that the total angular momentum has
the form of
\begin{equation}
\vec J = \vec L + \vec S + \vec I + \vec C ,
\end{equation}
which is the extension of
the conventional total angular momentum $\vec J = \vec L + \vec S$. The intrinsic
principal number $n_m$ denotes the intrinsic spatial dimension or radial quantization:
$n_c = 3$ represents strong interactions as an $SU(3)_C$ gauge theory. For QCD as the
$SU(3)_C$ gauge theory, there are nine gluons ($n_c^2 = 3^2 = 9$), which consist of
one singlet gluon $A_0$ with $c=0$, three degenerate gluons $A_1 \sim A_3$ with $c=1$,
and five degenerate gluons $A_4 \sim A_8$ with $c=2$; for QND as the $SU(2)_N \times
U(1)_Z$ gauge theory, one singlet gluon $A_0$ with $c=0$, three gluons $A_1 \sim A_3$
with $c=1$, and one gauge boson $A_8$ with $c=2$ are required. One explicit evidence
of colorspin and isospin angular momenta is strong isospin symmetry in nucleons, which
is postulated as the combination symmetry of colorspin and weak isospin in this
scheme. Another evidence is the nuclear magnetic dipole moment: the Lande spin
g-factors of the proton and neutron are respectively $g_s^p = 5.59$ and $g_s^n = -
3.83$, which are shifted from $2$ and $0$, because of contributions from colorspin and
isospin degrees of freedom as well as spin degrees of freedom. The mass ratio of the
proton and the constituent quark, $m_p/m_q \sim 2.79$, thus represents three intrinsic
degrees of freedom of color, isospin, and spin. In fact, the extrinsic angular
momentum associated with the intrinsic angular momentum may be decomposed by $\vec L =
\vec L_i + \vec L_c + \vec L_s$ where $\vec L_i$ is the angular momentum originated
from the isospin charge, $\vec L_c$ is the angular momentum originated from the color
charge, and $\vec L_s$ is the angular momentum originated from the spin charge. This
is supported by the fact that the orbital angular momentum $l_c$ of the nucleon has
the different origin from the color charge with the orbital angular momentum $l_i$ of
the electron from the isospin charge since two angular momenta have opposite
directions from the information of spin-orbit couplings in nucleus and atoms.
Extrinsic angular momenta have extrinsic parity $(-1)^l = (-1)^{(l_c + l_i + l_s)}$,
intrinsic angular momenta have intrinsic parity $(-1)^{(c + i + s)}$, and the total
parity becomes $(-1)^{(l + c + i + s)}$ for electric moments while extrinsic angular
momenta have extrinsic parity $(-1)^{(l + 1)} = (-1)^{(l_c + l_i + l_s + 1)}$,
intrinsic angular momenta have intrinsic parity $(-1)^{(c + i + s + 1)}$, and the
total parity becomes $(-1)^{(l + c + i + s + 1)}$ for magnetic moments.

Fermions increase their masses by decreasing their intrinsic
principal quantum numbers from the higher ones at higher energies to the lower ones at
lower energies. The coupling constant $\alpha_s$ of a non-Abelian gauge theory is
strong for the small $N_{sd}$ and is weak for the large $N_{sd}$ according to the
renormalization group analysis. The vacuum energy is described by the zero-point
energy in the unit of $\omega/2$ with the maximum number $N_{sd} \simeq 10^{61}$ and
the vacuum is filled with fermion pairs of up and down colorspins, isospins, or spins,
whose pairs behave like bosons quantized by the unit of $\omega$: this is analogous to
the superconducting state of fermion pairs. The intrinsic particle number $N_{sp}
\simeq 10^{30}$ (or $B \simeq 10^{-12}$, $L \simeq 10^{-9}$) characterizes
gravitational interactions for fermions with the mass $10^{-12}$ GeV, $N_{sp} \simeq
10^{6}$ (or $L_e \simeq 1$) characterizes weak interactions for electrons, and $N_{sp}
\simeq 1$ (or $B \simeq 1$) characterizes strong interactions for nucleons:
according to (\ref{gafe5}), $m_h = 0.94$ GeV, $c_f = 1/3$, $\alpha_s = 0.48$, $N_{sd} = 1$,
$M_G = 0.27$ GeV for a nucleon are realized.
Fundamental particles known as leptons and quarks are hence postulated as composite
particles with the color, isospin, and spin quantum numbers; quark is color triplet
state but lepton is color singlet. Note that if $N_{sp} > 1$ (or $B < 1$), it
represents a pointlike fermion and if $N_{sp} < 1$ (or $B > 1$), it represents a
composite fermion.

\subsection{$\Theta$ Constant and Quantum Numbers}

The invariance of gauge transformation provides
$\psi [\hat O_\nu] = e^{i \nu \Theta} \psi [\hat O]$ for the fermion wave function $\psi$ with the transformation of an operator $\hat O$ by
the class $\nu$ gauge transformation, $\hat O_\nu$:
the vacuum state characterized by the constant $\Theta$ is called the $\Theta$ vacuum \cite{Hoof2}.
The true vacuum is the superposition of all the $|\nu \rangle$ vacua with the phase $e^{i \nu \Theta}$:
$|\Theta \rangle = \sum_\nu e^{i \nu \Theta} |\nu \rangle$.
The topological winding number $\nu$ or the topological charge $q_s$ is defined by
\begin{equation}
\label{tonu}
\nu = \nu_+ - \nu_- = \int \frac{c_f g_s^2}{16 \pi^2} Tr G^{\mu \nu} \tilde G_{\mu \nu} d^4 x
\end{equation}
where the subscripts $+$ and $-$ denote moving axial vector and vector particles respectively in the presence of the gauge fields \cite{Atiy}.
The matter energy density generated by the surface effect is postulated by
\begin{equation}
\label{toma}
\rho_{m} \simeq \rho_c \simeq \frac{c_f g_s^2}{16 \pi^2} Tr G^{\mu
\nu} \tilde G_{\mu \nu} \simeq 10^{-47} \ \textup{GeV}^4
\end{equation}
which implies that the fermion mass is generated by the difference of vector (scalar)
and axial vector (pseudo-scalar) fermion numbers. In this aspect, the difference
number $N_{sd}$, the singlet fermion number $N_{ss}$, and the condensed singlet
fermion number $N_{sc}$ in intrinsic two-space dimensions respectively correspond to
$\nu$, $\nu_+$, and $\nu_-$ in three-space and one-time dimensions.  In the presence of the
$\Theta$ term, the singlet axial current is not conserved due to an Adler-Bell-Jackiw
anomaly \cite{Adle}:
\begin{equation}
\partial_\mu J_\mu^5 = \frac{N_f c_f g_s^2}{16 \pi^2} Tr G^{\mu \nu} \tilde G_{\mu \nu}
\end{equation}
with the flavor number of fermions $N_f$ and this reflects
degenerated multiple vacua. This illustrates mass generation by
the surface effect due to the field configurations with parallel
color electric and magnetic fields.
If $\nu = \rho_m/\rho_G$ is introduced from (\ref{tonu}) and (\ref{toma}),
a condition $\Theta \nu = 10^{-61}$ is satisfied and is compatible with the flat universe
condition $\Omega = 1 - 10^{-61}$.
The $\Theta$ value parameterized by $\Theta = 10^{-61} \rho_G/\rho_m$ is consistent with the observed
result, $\Theta < 10^{-9}$ in the electric dipole moment of the
neutron \cite{Alta}. The condition $\Theta \nu = 10^{-61}$ is related to the instanton mechanism
represented by the tiny tunneling amplitude $e^{-S}$
with the Euclidean action $S = \Theta \nu = 10^{-61}$ in the Euclidean spacetime.

The topological winding number $\nu$ is related to the intrinsic quantum number $n_m$ by $\nu = 1/n_m^{8}$.
The intrinsic principal number $n_m$ is also connected with $N_{sp}$ and $N_{sd}$:
$n_m^2 = N_{sp}$, $N_{sp}^2 = N_{sd}$, and $N_{sp}^{4} = 1/\nu$.
The relation between the intrinsic radius and the intrinsic quantum number might be ascribed by
\begin{math}
r_i = r_{0i} / n_m^2
\end{math}
with the radius $r_{0i} = 1/2 m_f \alpha_z \simeq N_{sp}/M_G$. Intrinsic
quantum numbers are exactly analogous to extrinsic quantum
numbers. The extrinsic principal number $n$ for the nucleon is
related to the nuclear mass number $A$ or the baryon number $B >
1$: $n^2 = A^{1/3}$, $n^4 = A^{2/3}$, $n^6 = B = A$. The relation
between the nuclear radius and the extrinsic quantum number is
outlined by
\begin{equation}
r = r_0 A^{1/3} = r_0 n^2
\end{equation}
with the radius $r_0 \approx 1.2$ fm and the nuclear principal number
$n$. This is analogous to the atomic radius $r_e = r_0 n_e^2$ with
the atomic radius $r_0$ and the electric principal number $n_e$:
the atomic radius $r_0 = 1/2 m_e \alpha_y$ is almost the same with
the Bohr radius $a_B = 1/m_e \alpha_e = 0.5 \times 10^{-8}$ cm.
These concepts are related to the constant nuclear density $n_B =
3/4 \pi r_0^3 = 1.95 \times 10^{38} \ \textup{cm}^{-3}$ or
Avogadro number $N_A = 6.02 \times 10^{23} \ \textup{mol}^{-1}$
and to the constant electron density $n_e = 3/4 \pi r_e^3 = 6.02
\times 10^{23} Z \rho_m/A$ with the matter energy density $\rho_m = \rho_B$ in the unit of
$\textup{g/cm}^3$ where the possible relation is $r_e = r_0
L^{1/3} = r_0 n_e^2$ with the lepton number $L$.

The $\Theta$ values according to (\ref{thvu}) become $\Theta_{Pl} \approx 10^{61}$,
$\Theta_{EW} \approx 10^{-4}$, $\Theta_{QCD} \approx 10^{-12}$, and $\Theta_{0}
\approx 10^{-61}$ at different stages. The scope of $\Theta = 10^{61} \sim 10^{-61}$
corresponds to the scope of $\nu = 10^{-122} \sim 10^{0}$ to satisfy the flat universe
condition $\nu \Theta = 10^{-61}$: the maximum quantization number $N_{sp} \simeq N_R
\simeq 10^{30}$ and $N_G \simeq 4 \pi N_R^3/3 \simeq 10^{91}$. The maximum wavevector mode
$N_R = (\rho_G/\Theta \rho_B)^{1/2} = 10^{30}$ of the QCD vacuum is obtained. These
describe possible dualities between intrinsic quantum numbers and extrinsic quantum
numbers: $n_m$ and $n$, $N_{sp}$ and $A^{1/3}$, and $1/\nu$ and $A^{4/3}$ for
baryons.

Baryon mass generation from the vacuum is described by $\rho_B
\equiv \Omega_B \rho_c \simeq 10^{-61} \Omega_B \rho_G/\Theta$ at the
strong scale. This $\Theta$ term as the surface term modifies the
original QCD for strong interactions \cite{Frit}, which has quark
mass term violating gauge invariance, and suggests the mass
generation as the nonperturbative breaking of gauge and chiral
invariance through DSSB.

\section{Conclusions}

This paper claims that the longstanding problems of QCD as an $SU(3)_C$ gauge theory,
the confinement mechanism and $\Theta$ vacuum, can be resolved and quantum
nucleardynamics (QND) is produced in terms of dynamical spontaneous symmetry breaking
(DSSB) through the condensation of singlet gluons. DSSB consists of two simultaneous
mechanisms; the first mechanism is the explicit symmetry breaking of gauge symmetry,
which is represented by the color factor $c_f$ and the strong coupling constant $g_s$,
and the second mechanism is the spontaneous symmetry breaking of gauge fields, which
is represented by the condensation of singlet gauge fields. QCD produces QND as an
$SU(2)_N \times U(1)_Z$ gauge theory through DSSB. The combination of the confinement
mechanism and $\Theta$ vacuum suggests the DSSB mechanism in QCD analogous to the
Higgs mechanism in the Glashow-Weinberg-Salam (GWS) model as the $SU(2)_L \times
U(1)_Y$ electroweak theory. The DSSB of local gauge symmetry and global chiral
symmetry triggers the axial current anomaly. P violation and C violation are maximal
but CP violation and T violation are minimal during DSSB; explicit evidence is the
observation of pseudoscalar and vector mesons without their parity partners and the
small $\Theta$ value $(\leq 10^{-9})$ for the neutron electric dipole moment. CVC and
PCAC in strong interactions are well explained through the condensation of singlet
gluons. The $\Theta$ constant is parameterized by $\Theta = 10^{-61} \ \rho_G/\rho_m$
with the vacuum energy density $\rho_G = M_G^4$ and the matter energy density
$\rho_m$. Massless gauge bosons (photons) as Nambu-Goldstone (NG) bosons indicate the
DSSB mechanism of local gauge symmetry and global chiral symmetry. Quantized gauge
bosons, i.e., gluons, are massive and yield the Yukawa potential; this is understood
as the confinement mechanism of massive gluons limited to relatively short range
propagation. Phase transition occurs when singlet gluons, rather than other scalar
bosons, acquire vacuum expectation values and reduce the masses of gauge bosons: the
gauge boson mass square is $M^2_G = M^2_{H} - c_f g_s^2 \langle \phi \rangle^2 = c_f
g_s^2 [A_{0}^2 - \langle \phi \rangle^2]$ with the grand unification mass $M_H$, the
strong coupling constant $g_s$, and the singlet gluon condensation $\langle \phi
\rangle$.

QND as an $SU(2)_N \times U(1)_Z$ gauge theory is the analogous dynamics of the GWS model
as an $SU(2)_L \times U(1)_Y$ gauge theory for weak interactions. The confinement mechanism
can be reinterpreted in terms of the analogy property; the phenomenology of strong
force provides the analogous characteristics with weak force. At short range, the
Coulomb type repulsive source is dominant but at long range, the Yukawa potential is
dominant due to massive gluons; the effective coupling constant $G_R/\sqrt{2} = c_f
g_s^2 /8 M_G^2 \approx 10 \ \textup{GeV}^{-2}$ and color mixing angle $\sin^2 \theta_R
=1/4$ are predicted just as $G_F/\sqrt{2} = g_w^2 /8 M_W^2 \approx 10^{-5} \
\textup{GeV}^{-2}$ and $\sin^2 \theta_W = 1/4$ in weak interactions. The color
$SU(3)_C$ symmetry is spontaneously broken to the $SU(2)_N \times U(1)_Z$ symmetry and
then to the $U(1)_f$ symmetry. Strong coupling constants for baryons are $\alpha_b =
c_f^b \alpha_s = \alpha_s/3 $, $\alpha_n = c_f^n \alpha_s = \alpha_s/4$, $\alpha_z =
c_f^z \alpha_s = \alpha_s/12$, and $\alpha_f = c_f^f \alpha_s = \alpha_s/16$ as
symmetric color interactions and $- 2 \alpha_s/3 $, $- \alpha_s/2 $, $- \alpha_s/6$,
and $- \alpha_s/8$ as asymmetric color interactions: $c_f^n = \sin^2 \theta_R$ and
$c_f^f = \sin^4 \theta_R$. The color factors are $c^s_f = (c_f^b, c_f^n, c_f^z, c_f^f)
= (1/3, 1/4, 1/12, 1/16)$ for symmetric interactions and $c^a_f = (-2/3, -1/2, -1/6,
-1/8)$ for asymmetric interactions.  The symmetric charge factors reflect intrinsic even
parity with repulsive force while the asymmetric charge factors reflect intrinsic odd
parity with attractive force; this suggests electric-magnetic duality before DSSB.
Features of gauge theories, such as gauge invariance, vacuum problem, and discrete
symmetry breaking, for both strong and weak interactions are explained through DSSB
mechanism; the $U(1)_f$ gauge theory is proposed for massless gauge boson (photon)
interactions just as the $U(1)_e$ gauge theory is for photon interactions. Through
DSSB, QCD for strong interactions leads to QND as an $SU(2)_N \times U(1)_Z$ gauge theory,
which plays a dominant role in the low energy regime of strong
interactions. The confinement mechanism and $\Theta$ vacuum are simultaneously
resolved in terms of DSSB, which rigorously shows the confinement of quarks and gluons
inside the hadron and creates massless gauge bosons as NG bosons.

Mass generation mechanism for the hadron is suggested in terms of the DSSB of gauge
symmetry and chiral symmetry known as the dual pairing mechanism of the
superconducting state; the constituent quark model is thus justified as an effective
model of QCD at low energies. Dual pairing mechanism provides the hadron mass
generated from the vacuum: $M_G = \sqrt{\pi} m_h c_f \alpha_s \sqrt{N_{sd}}$ where the
difference number of axial-nonaxial singlet fermions $N_{sd} \simeq 1$ for the color
charge in the case of the proton mass $0.94$ GeV. $N_{sd} = N_{ss} - N_{sc}$ is
defined with the singlet particle number $N_{ss}$ and the condensed particle number
$N_{sc}$: electric-magnetic duality is closely related to quantum numbers $N_{sd}$ and
$N_{sc}$. The relation between the $\Theta$ constant and the difference number
$N_{sd}$ is given by $\Theta = \pi^2 m_h^4 c_f^4 \alpha_s^4 N_{sd}^2/10^{61} \rho_c$.
The difference number $N_{sd}$ in intrinsic two-space dimensions suggests the
introduction of a degenerated particle number $N_{sp}$ in the intrinsic radial
coordinate and an intrinsic principal number $n_m$; particle numbers are connected
with the relation $n_m^4 = N_{sp}^2 = N_{sd}$ and the Dirac quantization condition
$\sqrt{c_f} g_s g_{sm} = 2 \pi N_{sp}$ is satisfied. The intrinsic principal quantum
number $n_m$ consists of three quantum numbers, that is, $n_m = (n_c, n_i, n_s)$ where
$n_c$ is the intrinsic principal quantum number for the color space, $n_i$ is the
intrinsic principal quantum number for the isospin space, $n_s$ is the intrinsic
principal quantum number for the spin space. A baryon therefore possesses a set of
intrinsic quantum numbers $(n_c, n_i, n_s)$ to represent its intrinsic quantum states.
The concept automatically adopts the three types of intrinsic angular momentum
operators, $\hat C$, $\hat I$, and $\hat S$, when intrinsic potentials for colorspin,
isospin, and spin charges are central so that they depend on the intrinsic radial
distance: for instance, the color potential in strong interactions is dependent on the
radial distance. The principal number $n_m$ in intrinsic space quantization is very
much analogous to the principal number $n$ in extrinsic space quantization and the
intrinsic angular momenta are analogous to the extrinsic angular momentum so that the
total angular momentum has the form of $\vec J = \vec L + \vec S + \vec I + \vec C$,
which is the extension of the conventional total angular momentum $\vec J = \vec L +
\vec S$. One explicit evidence of colorspin and isospin angular momenta is strong
isospin symmetry in nucleons, which is postulated as the combination symmetry of
colorspin and weak isospin in this scheme. Another evidence is the nuclear magnetic
dipole moment: the Lande spin g-factors of the proton and neutron are respectively
$g_s^p = 5.59$ and $g_s^n = - 3.83$, which are shifted from $2$ and $0$, because of
contributions from colorspin and isospin degrees of freedom as well as spin degrees of
freedom. The mass ratio of the proton and the constituent quark, $m_p/m_q \sim 2.79$,
thus represents three intrinsic degrees of freedom of colorspin, isospin, and spin. In
fact, the extrinsic angular momentum may be decomposed by $\vec L = \vec L_i + \vec
L_c + \vec L_s$ where $\vec L_i$ is the angular momentum originated from the isospin
charge, $\vec L_c$ is the angular momentum originated from the color charge, and $\vec
L_s$ is the angular momentum originated from the spin charge. Fermions increase their
masses by decreasing their intrinsic principal quantum numbers from the higher ones at
higher energies to the lower ones at lower energies. The coupling constant $\alpha_s$
is strong for the small $N_{sd}$ and is weak for the large $N_{sd}$ according to the
renormalization group analysis. The vacuum energy is described by the zero-point
energy in the unit of $\omega/2$ with the maximum number $N_{sd} \simeq 10^{61}$ and
the vacuum is filled with baryon pairs of up and down colorspins, isospins, or spins,
whose pairs behave like bosons quantized by the unit of $\omega$: this is analogous to
the superconducting state of fermion pairs.

The invariance of gauge transformation provides $\psi [\hat O_\nu] = e^{i \nu \Theta}
\psi [\hat O]$ for the fermion wave function $\psi$ with the transformation of an
operator $\hat O$ by the class $\nu$ gauge transformation, $\hat O_\nu$: the vacuum
state characterized by the constant $\Theta$ is called the $\Theta$ vacuum. The true
vacuum is the superposition of all the $|\nu \rangle$ vacua with the phase $e^{i \nu
\Theta}$: $|\Theta \rangle = \sum_\nu e^{i \nu \Theta} |\nu \rangle$. The topological winding number
$\nu$ or the axial charge $q_5$ at the strong scale is defined by $\nu = \int
\frac{c_f g_s^2}{16 \pi^2} Tr G^{\mu \nu} \tilde G_{\mu \nu} d^4 x$ where the matter
density due to the surface effect is also defined by $\rho_{m} = \frac{c_f g_s^2}{16
\pi^2} Tr G^{\mu \nu} \tilde G_{\mu \nu}$. In the presence of the $\Theta$ term, the
singlet axial current is not conserved due to an anomaly: $\partial_\mu J_\mu^5 =
\frac{N_f c_f g_s^2}{16 \pi^2} Tr G^{\mu \nu} \tilde G_{\mu \nu}$ with the flavor
number of fermions $N_f$ and it leads to degenerated multiple vacuum. The $\Theta$
value parameterized by $\Theta = 10^{-61} \rho_G/\rho_m$ is consistent with the observed
results, $\Theta < 10^{-9}$ in the electric dipole moment of the neutron. The
topological winding number $\nu = \rho_m/\rho_G$ is related to the intrinsic quantum number $n_m$ by
$\nu = 1/n_m^{8}$. The intrinsic principal number $n_m$ is also connected with
$N_{sp}$ and $N_{sd}$: $n_m^2 = N_{sp}$, $N_{sp}^2 = N_{sd}$, and $N_{sp}^{4} =
1/\nu$. Intrinsic quantum numbers are exactly analogous to extrinsic quantum
numbers. The extrinsic principal number $n$ for the nucleon is related to the nuclear
mass number $A$ or the baryon quantum number $B>1$: $n^2 = A^{1/3}$, $n^4 = A^{2/3}$,
$n^6 = B = A$. The relation between the nuclear radius and the extrinsic quantum
number is outlined by $r = r_0 A^{1/3} = r_0 n^2$ with the radius $r_0 \approx 1.2$ fm
and the nuclear principal number $n$ and is analogous to the atomic radius $r_e = a_0
n_e^2$ with  the atomic radius $a_0 = 1/2 m_e \alpha_y$ or the Bohr radius $a_B =
1/m_e \alpha_e = 0.5 \times 10^{-8}$ cm and the electric principal number $n_e$. These
concepts are related to the constant nuclear density $n_B = 1.95 \times 10^{38} \
\textup{cm}^{-3}$ and Avogadro number $N_A = 6.02 \times 10^{23} \ \textup{mol}^{-1}$.
The maximum wavevector mode $N_R = (\rho_G/\Theta \rho_B)^{1/2} = 10^{30}$ of the QCD
vacuum is obtained. The $\Theta$ term as the surface term modifies the original QCD
for strong interactions, which has the fermion mass problem violating gauge
invariance, and suggests mass generation as the nonperturbative breaking of gauge
and chiral invariance through DSSB.

Significant consequences of this work are summarized in the following. QCD as an
$SU(3)_C$ gauge symmetry generates QND as an $SU(2)_N \times U(1)_Z$ gauge theory or a
$U(1)_f$ gauge theory through DSSB induced by the condensation of singlet gluons. The
quantization of vacuum and matter energies is suggested by conserved particle numbers.
Massive gluons make confinement and massless gauge bosons indicate the DSSB mechanism
of local gauge symmetry and global chiral symmetry, which initiates the axial current
anomaly. The nonperturbative solution of QCD in the low energy limit is explicitly
obtained from DSSB; confinement-deconfinement phase transition is derived and the
mechanism of hadron formation can be given. The existence of the dual Meissner effect
in analogy with superconductivity phase transition provides a deeper understanding of
particle bound mechanism, which is described both by the gluon condensation as the
confinement mechanism and by massless gauge boson interactions as NG bosons of strong
confinement. The mechanism of hadron mass generation is suggested in terms of the DSSB
of gauge symmetry and chiral symmetry; hadron mass generation mechanism is compatible
with the constituent quark model as an effective model of QCD at low energies. The
$\Theta$ vacuum or strong CP problem relevant for the axial current anomaly is
resolved in terms of DSSB; the vacuum condensation is the source of discrete symmetry
breaking. Hadron mass generation and $\Theta$ vacuum are thus
understood in terms of intrinsic and extrinsic quantum numbers. The analogy property
emphasizes that strong force is the analogous partner of weak force and that massless
gauge bosons mediate electromagnetic interactions. This proposal may provide a turning
point toward the understanding of the confinement through DSSB at relatively low
energies.

\end{document}